\documentclass[11pt]{article}
\pdfoutput=1
\usepackage{jheppub}
\usepackage[usenames,dvipsnames,table]{xcolor}
\usepackage{graphicx,amsmath,amssymb,amsthm,multirow,array,bm,bbm,esint,physics}
\usepackage[mathscr]{eucal}
\usepackage[bbgreekl]{mathbbol}
\usepackage{epsf,amsfonts}
\usepackage{slashed}
\usepackage[numbers,sort&compress]{natbib}
\usepackage{minibox}
\usepackage{pdflscape}
\usepackage{comment}
\usepackage{tikz,pgf}

\DeclareSymbolFontAlphabet{\mathbb}{AMSb}%

\usepackage[dvipsnames]{xcolor}

\usepackage{slashed}

\def \str{\mbox{str\,}}

\def\+{{+\!\!\!+}}		%+ light-cone

\newcommand{\cD}{\mathcal{D} }

\newcommand{\cK}{\mathcal{K} }
\newcommand{\cL}{\mathcal{L} }
\newcommand{\cM}{\mathcal{M} }
\newcommand{\cN}{\mathcal{N} }
\newcommand{\cO}{\mathcal{O} }
\newcommand{\cP}{\mathcal{P} }

\newcommand{\cR}{\mathcal{R} }

\newcommand{\cW}{\mathcal{W} }

\newcommand{\bA}{\mathbf{A} }

\usepackage{bbold}
\usepackage{cleveref}
\usepackage{tensor}

\newcommand{\bb}[1]{\mathbb{#1}}

\newcommand{\sbk}[1]{\left[ #1\right]}
\newcommand{\cbk}[1]{\left\{#1\right\}}

\newcommand{\tp}{\theta}       %   theta
\newcommand{\tm}{\bar{\tp}}  % barred eta
\newcommand{\etp}{\eta}       %  eta
\newcommand{\etm}{\bar{\etp}}  % barred eta

\newcommand{\z}{z}
\newcommand{\w}{w}
\newcommand{\zb}{\bar{z}}     % z bar
\newcommand{\wb}{\bar{w}}   % w bar

\newcommand{\bth}{\vartheta} % boundary theta
\newcommand{\btp}{\bbtheta}    % bold theta
\newcommand{\btm}{\bar{\btp}}  % barred bold theta
\newcommand{\bep}{\bbeta}    % bold eta
\newcommand{\bem}{\bar{\bep}}  % barred  bold eta

\newcommand{\bz}{\mathbb{z}}   % bold z
\newcommand{\bzb}{\bar{\bz}}   % barred bold z
\newcommand{\bw}{\mathbb{w}} % bold w
\newcommand{\bwb}{\bar{\bw}}  % barred bold w

\newcommand{\del}{\partial}   % partial
\newcommand{\delb}{\bar{\del}}  %partial bar

\newcommand{\dz}{\partial_{\z}}  % partial z
\newcommand{\dzb}{\partial_{\bar{\z}}}  % partial z bar
\newcommand{\dw}{\partial_{\w}}  % partial w
\newcommand{\dwb}{\partial_{\bar{\w}}}  % partial w bar

\newcommand{\dtp}{\del_{\tp}}   % partial theta
\newcommand{\dtm}{\del_{{\bar{\tp}}}}     % partial theta bar
\newcommand{\dep}{\del_{\etp}}   % partial theta
\newcommand{\dem}{\del_{{\bar{\etp}}}}     % partial theta bar

\newcommand{\Dpn}{\mathsf{D}}   % serid D
\newcommand{\Dmn}{\bar{\mathsf{D}}}  % serif D bar

\newcommand{\Dp}{D}   % D
\newcommand{\Dm}{\bar{D}}  % D bar

\newcommand{\bDp}{\mathbb{D}}
\newcommand{\bDm}{\bar{\mathbb{D}}}
\newcommand{\bD}{\mathcal{D}}

\newcommand{\bdel}{\reflectbox{$\mathbb{6}$}}   % bold partial
\newcommand{\bdelb}{\bar{\bdel}}  %bold partial bar
\newcommand{\we}{\Omega} %conformal factor
\newcommand{\swe}{\Sigma}   %super conformal factor
\newcommand{\sfr}{\mathcal{E}} %super zweibein
\newcommand{\nsfr}{\widetilde{\mathcal{E}}} %super zweibein

\newcommand{\rap}{\Lambda} % bosonic rapidity/boost
\newcommand{\srap}{\mathsf{\Lambda}} %super rapidity/boost

\newcommand{\spc}{\varphi} %spin connection
\newcommand{\lgen}{\mathbb{L}} % Lorentz generator

\newcommand{\tangen}{T} %super tangential vector
\newcommand{\normal}{N} %super normal vector

\newcommand{\ie}{\textit{i.e.}, }
\newcommand{\eg}{\textit{e.g.}, }

\newcommand{\ads}{\mathrm{AdS} }
\newcommand{\sch}{\mathrm{Sch} }
\newcommand{\ssch}{\mathrm{sSch} }

\title{Gravitational Edge Mode in $\cN=1$ Jackiw-Teitelboim Supergravity}

\author[\, a,b]{Kyungsun Lee}
\author[\, c]{Akhil Sivakumar}
\author[\, a,c,d]{Junggi Yoon}
\affiliation[a]{School of Physics, Korea Institute for Advanced Study\\
85 Hoegiro Dongdaemun-gu, Seoul 02455, Korea}
\affiliation[b]{School of Physics and Chemistry, Gwangju Institute of Science and Technology,\\
123 Cheomdan- gwagiro, Gwangju 61005, Korea}
\affiliation[c]{Asia Pacific Center for Theoretical Physics,\\77 Cheongam-ro, Nam-gu, Pohang-si, Gyeongsangbuk-do, 37673, Korea}
\affiliation[d]{Department of Physics, POSTECH\\ 77 Cheongam-ro, Nam-gu, Pohang-si, Gyeongsangbuk-do, 37673, Korea}

\emailAdd{kyungsun.cogito.lee@gmail.com}
\emailAdd{akhil.sivakumar@apctp.org}
\emailAdd{junggiyoon@gmail.com}

\abstract{ We study the gravitational edge mode in the $\cN=1$ Jackiw-Teitelboim~(JT) supergravity on the disk and it $osp(2|1)$ BF formulation. We revisit the derivation of the finite-temperature Schwarzian action in the conformal gauge of the bosonic JT gravity through wiggling boundary and the frame fluctuation descriptions. Extending our method to $\cN=1$ JT supergravity, we derive the finite-temperature super-Schwarzian action for the edge mode from both the wiggling boundary and the superframe field fluctuation. We emphasize the crucial role of the supersymmetric version of the inversion formula in elucidating the relation between the isometry and the $OSp(2|1)$ gauging of the super-Schwarzian action. In $osp(2|1)$ BF formulation, we discuss the asymptotic AdS condition. We employ the Iwasawa-like decomposition of $OSp(2|1)$ group element to derive the super-Schwarzian action at finite temperature. We demonstrate that the $OSp(2|1)$ gauging arises from inherent redundancy in the Iwasawa-like decomposition. We also discuss the path integral measure obtained from the Haar measure of $OSp(2|1)$.
}

\begin{document}
\maketitle

%%%%%%%%%%%%%%%%%%%%%%%%%%%%%%%%%%%%%%%%%%%%%%%%%%%%%%%%%%%
%%%%%%%%%%%%%%%%%%%%%%%%%%%%%%%%%%%%%%%%%%%%%%%%%%%%%%%%%%%
\section{Introduction}
\label{sec: introduction}
%%%%%%%%%%%%%%%%%%%%%%%%%%%%%%%%%%%%%%%%%%%%%%%%%%%%%%%%%%%
%%%%%%%%%%%%%%%%%%%%%%%%%%%%%%%%%%%%%%%%%%%%%%%%%%%%%%%%%%%

The exploration of quantum gravity remains one of the most profound challenges in theoretical physics. Among the various approaches to quantum gravity, lower-dimensional gravity models have provided insightful playgrounds for understanding of quantum gravity, holography, and the structure of spacetime. The two-dimensional Jackiw-Teitelboim (JT) gravity model~\cite{Almheiri:2014cka,Jensen:2016pah,Maldacena:2016upp,Cvetic:2016eiv,Das:2017pif,Mandal:2017thl,Goel:2020yxl,Joung:2023doq,Bak:2023zkk} has emerged as a pivotal framework, offering profound insights into the quantum aspects of black holes, holography and the nature of gravitational dynamics. The supersymmetric extension\footnote{See Ref.~\cite{Chamseddine:1991fg,Chamseddine:1991qu} for the JT supergravity in the context of superstring.} of the JT gravity for the nearly-AdS$_2$~\cite{Forste:2017kwy,Forste:2017apw,Cardenas:2018krd,Forste:2020xwx,Fan:2021wsb} has been spotlighted in the study of the microstates of BPS black holes~\cite{Heydeman:2020hhw,Boruch:2022tno,Iliesiu:2022kny,Ciceri:2023mjl}. The JT supergravity further enriches the theoretical landscape, facilitating a deep exploration of the interplay between supergravity and supersymmetric quantum mechanics~\cite{Fu:2016vas,Narayan:2017hvh,Peng:2017spg,Kanazawa:2017dpd,Yoon:2017gut,Bulycheva:2018qcp,Berkooz:2020xne,Heydeman:2022lse,Turiaci:2023jfa}.

The dynamics of the JT gravity can be captured by the wiggling boundary of $AdS_2$~\cite{Maldacena:2016upp}, or equivalently by the boundary fluctuation of the metric. As observed in the fractional quantum Hall effect~\cite{Wen:1989mw,Wen:1990qp,Wen:1990se,Stone:1990by}, the broken radial diffeomorphism on the boundary of AdS spacetime is responsible for these boundary degrees of freedom, called as the \textit{would-be gauge mode} or the \textit{gravitational edge mode mode}. The investigation on the edge mode has been carried out in various gravity context~\cite{Balachandran:1994up,Balachandran:1995qa,Arcioni:2002vv,Carlip:1995cd,Takayanagi:2019tvn,Donnelly:2020teo,David:2022jfd,Mertens:2022ujr,Wong:2022eiu,Joung:2023doq,Mukherjee:2023ihb}. Recently, the gravitational edge mode in the bosonic JT gravity has been revisited by considering the mapping between base AdS and target AdS~\cite{Joung:2023doq,Ferrari:2024kpz}. Here the inversion formula play a crucial role both in the derivation of Schwarzian action and in elucidating its $PSL(2,\mathbb{R})$ gauging\cite{Joung:2023doq}.

In this paper, we aim at extending the discussion of the gravitational edge mode into the $\cN=1$ JT supergravity. This paper is organized as follows. In Section~\ref{sec: review: gravitational edge mode in bosonic JT gravity}, we review the gravitational edge modes in the bosonic JT gravity, emphasizing the derivation of the Schwarzian action from the perspectives of wiggling boundary description and frame field fluctuation in the conformal gauge. This section sets the stage for our exploration by detailing the conformal gauge, providing a solid groundwork for the supersymmetric extensions that follow. In Section~\ref{sec: super jt gravity}, we extend the derivation to the $\cN=1$ JT supergravity. We revisit the superconformal gauge and an appropriate boundary term, employing both wiggling boundary descriptions and superframe fluctuations to derive the super-Schwarzian action for gravitational edge modes. We elucidate the origin of the $OSp(2|1)$ gauging of the super-Schwarzian action, emphasizing the importance of the supersymmetric version of the inversion formula. In Section~\ref{sec: bf description}, we study the $osp(2|1)$ BF formulation of JT supergravity, where we discuss the asymptotic AdS condition as a constraint of BF theory. We employ the Iwasawa-like decomposition, apt for capturing the monodromy of disk topology, to derive the super-Schwarzian action. We demonstrate the redundancy intrinsic to the Iwasawa-like decomposition results in the $OSp(2|1)$ gauging of the super-Schwarzian action. Additionally, we discuss the path integral measure obtained from the Haar measure of the $OSp(2|1)$ group. In Section~\ref{sec: conclusion}, we make concluding remarks.

\begin{comment}
\begin{center}
	
\begin{tabular}{|c|c|}
	\hline
	symbol & macro\\
	\hline 
	$\tp$, $\tm$ & $\backslash$tp,$\backslash$tm\\
	$\etp$,  $\etm$  & $\backslash$etp, $\backslash$etm \\ 
	$\z$,  $\zb$  & $\backslash$z, $\backslash$zb  \\ 
	$\w$,  $\wb$  & $\backslash$w, $\backslash$wb  \\ 
	$\btp$,  $\btm$  & $\backslash$btp, $\backslash$btm  \\ 
	$\bep$,  $\bem$  & $\backslash$bep, $ \backslash $bem  \\ 
	$\bz$,  $\bzb$  & $\backslash$bz, $\backslash$bzb   \\ 
	$\bw$,  $\bwb$  & $\backslash$bw, $\backslash$bwb  \\ 
	$\del$,  $\delb$,	$\bdel$,  $\bdelb$  & $\backslash$del, $\backslash$delb, $\backslash$bdel, $\backslash$bdelb  \\ 
	$\dz$,  $\dzb$  & $\backslash$dz, $\backslash$dzb  \\ 
	$\dw$,  $\dwb$  & $\backslash$dz, $\backslash$dwb  \\ 
	$\dtp$,  $\dtm$  & $\backslash$dtp, $\backslash$dtm  \\ 
	$\dep$,  $\dem$  & $\backslash$dep, $\backslash$dem \\ 
	$\Dpn$,  $\Dmn$  & $\backslash$Dpn, $\backslash$Dmn   \\ 
	$\Dp$,  $\Dm$  & $\backslash$Dp, $\backslash$Dm  \\ 
	$\bDp$,  $\bDm$, $\bD$  & $\backslash$bDp, $\backslash$bDm, $\backslash$bD  \\ 
	$\we$,  $\swe$  & $\backslash$we, $\backslash$swe  \\ 
	$\sfr$,  $\nsfr$  & $\backslash$sfr, $\backslash$nsfr \\ 
	$\rap$,  $\srap$  & $\backslash$rap, $\backslash$srap  \\ 
 $\bth$,  $\bD$  & $\backslash$bth, $\backslash$bD  \\ 
\end{tabular}

\end{center}
\end{comment}

%%%%%%%%%%%%%%%%%%%%%%%%%%%%%%%%%%%%%%%%%%%%%%%%%%%%%%%%%%%
%%%%%%%%%%%%%%%%%%%%%%%%%%%%%%%%%%%%%%%%%%%%%%%%%%%%%%%%%%%
\section{Review: Gravitational Edge Mode in Bosonic JT Gravity}
\label{sec: review: gravitational edge mode in bosonic JT gravity}
%%%%%%%%%%%%%%%%%%%%%%%%%%%%%%%%%%%%%%%%%%%%%%%%%%%%%%%%%%%
%%%%%%%%%%%%%%%%%%%%%%%%%%%%%%%%%%%%%%%%%%%%%%%%%%%%%%%%%%%

We begin by reviewing the gravitational edge mode in the bosonic JT gravity~\cite{Maldacena:2016upp,Joung:2023doq}. The action for JT gravity is given by 
\begin{equation}
	I_{\text{JT}}\,=\, \int_{\mathcal{M}} \dd^2 x\sqrt{g}\,\Phi(R+2)+ 2\int_{\partial\mathcal{M}}\dd u \sqrt{h}\,\Phi(K-K_{0}) , \label{eq: jt action}
\end{equation}
where $\mathcal{M}$ denotes the two-dimensional Euclidean AdS space. The coordinate for the boundary $\partial\cM$ is denoted by $u$. We add the extrinsic curvature $K$ on the boundary of $\ads_2$ for the well-defined variational principle. The counterterm $K_0$ is chosen to be the value of the extrinsic curvature for the exact $\ads_2$ metric. The fields $\Phi$, $g$ and $h$ denotes the dilaton, the bulk metric and the induced boundary metric, respectively.

%%%%%%%%%%%%%%%%%%%%%%%%%%%%%%%%%%%%%%%%%%%%%%%%%%%%%%%%%%%
\subsection{Conformal Gauge}
\label{sec: conformal gauge}
%%%%%%%%%%%%%%%%%%%%%%%%%%%%%%%%%%%%%%%%%%%%%%%%%%%%%%%%%%%

The bulk equations of motion of JT gravity~\eqref{eq: jt action} are 
\begin{equation}
		R+2\,=\,0 \ , \qquad 
		\left(\nabla_{m}\nabla_{n}-g_{mn}\,\nabla^{2}+g_{mn}\right)\Phi\,=\,0 \ . \label{eq: bosonic JT eom}
\end{equation}
One of the classical solutions to this dilaton-gravity equations~\eqref{eq: bosonic JT eom} in the conformal gauge can be found to be\footnote{The change of coordinates by $ w= r e^{i \tau}$   with 	$\tau =  y + \frac{\pi}{2}  , \,  r = \left[\frac{\rho-1}{\rho+1}\right]^{\frac{1}{2}} $ gives the metric of the Euclidean black hole with inverse temperature $\beta = 2 \pi $:
\begin{equation}
	\dd s^{2} \,=\, (\rho^{2}-1) \dd y^{2} + \frac{\dd \rho^{2}}{\rho^{2} - 1} \ , \qquad \phi \,=\,  \rho \ .
\end{equation}
}
\begin{equation}\label{eq:BosBG}
		\dd s^{2} = 4e^{2 \Omega} \dd \w \dd \wb\;\; , \quad e^{2 \Omega} = \frac{1\, }{\left(1- \w \wb \right)^{2}} \;\;, \quad  \Phi\, =\,  \frac{1+\w\wb}{1-\w\wb} \ .
\end{equation}
In the $\cN=1$ JT supergravity, we will consider a superconformal gauge with disk topology. Hence, it is useful to review the gravitational edge mode from the conformal gauge in the disk topology, \ie Poincare disk. In this Poincare disk solution with periodicity $\log w \sim \log w + 2\pi i$, the tip of the Euclidean cigar, which is the analytic continuation of the horizon in the Lorentzian black hole, corresponds to $w \wb = 0$. The boundary is located at $w\wb  = 1$. 
 
The isometry of the Poincare disk is given by
\begin{align}
    w \;\;\longrightarrow \;\;  e^{i\phi}{ w + v\over \bar{v} w +1}\qquad \mbox{where} \;\; v\in \mathbb{C}\;,\;\; |v|<1\;,\;\; \phi\in \mathbb{R}\ .\label{eq: isometry of poincare disk 1}
\end{align}
This isometry can be written in a way that $PSU(1,1)$ is manifest.\footnote{
Using the map $z=  \frac{- i w+ 1 }{w -i}$ from the Poincare disk to the upper half plane, one can obtain the manifestly-$PSL(2,\mathbb{R})$ isometry of the upper half plane.
\begin{equation}
	z \;\;\longrightarrow \;\;  \frac{a z + b}{c z + d} \qquad (a d - b c =1 ) \ , 
\end{equation}
where $p =  \frac{1}{2}(a + d + i (b-c))$, $q = \frac{1}{2}(b + c +i (a -d))$.

} 
\begin{equation}
	w\;\;\longrightarrow\;\;  \frac{ p \, w + q}{\bar{q} w + \bar{p}}\qquad \mbox{where} \quad p\,,\;q\in \mathbb{C}\;\;,\;\;  |p|^{2} - |q |^{2} = 1 \ ,\label{eq: isometry of poincare disk 2}
\end{equation}
where the relation of the parameters is found to be
\begin{align}
    {p\over \bar{p}}\,=\, e^{i\phi}\;\;,\quad {q\over p}\,=\, v\ .
\end{align}
The dilaton solution is transformed under the $PSU(1,1)$ isometry:
\begin{equation}
\Phi\,=\, \frac{1+w\wb}{1-w\wb}\;\;\longrightarrow \;\; \Phi\,=\,   \frac{\left(|p|^{2}+|q|^{2} \right)(1+w\wb) + 2 p\bar{q} \, w + 2 \bar{p}q\,  \wb }{1-w \wb}  \ .
\end{equation}

%%%%%%%%%%%%%%%%%%%%%%%%%%%%%%%%%%%%%%%%%%%%%%%%%%%%%%%%%%%
\subsection{Wiggling Boundary in Conformal Gauge}
\label{sec: wiggling boundary in conformal gauge}
%%%%%%%%%%%%%%%%%%%%%%%%%%%%%%%%%%%%%%%%%%%%%%%%%%%%%%%%%%%

The wiggling boundary was introduced to account for the physical degrees of freedom on the boundary of JT gravity, which gives rise to the nearly-$\ads_2$ geometry~\cite{Maldacena:2016upp}. We will now review the derivation of the Schwarzian action from the dynamical wiggling boundary in the conformal gauge.

It was pointed out in \cite{Joung:2023doq} that the variation of the action with respect to the metric and dilaton leads to the boundary condition of the bulk metric and the bulk dilaton while the variation with respect to the wiggle boundary (\ie the boundary shape) yields the equation of motion of boundary physical degrees of freedom, namely the gravitational edge mode. Schematically, the variation of action is
\begin{align}
&\delta I_{\text{JT}} \,=\,  (E.o.M)\cr
&+\int_{\partial\mathcal{M}} \left(\cdots \right)\delta h|_{\partial\cM :\text{fixed}} + \int_{\partial\mathcal{M}}\left(\cdots \right) \delta{\phi}|_{\partial\cM :\text{fixed}} + \int_{\partial\mathcal{M}}\left(\cdots \right)\delta \text{(wiggling)}|_{h,\phi :\text{fixed}} \ .
\end{align}

The dynamical boundary described by Schwarzian action is wiggling close to the full $EAdS_2$ boundary $\sqrt{w\wb}=1-\epsilon$. Hence one can perturbatively parameterize the dynamical wiggling around the boundary of the disk by
\begin{equation}
\begin{aligned}
	\z  &\,=\, -ie^{i\tau(u)}\left(1-\epsilon\sigma(u)+\cO(\epsilon^2)\right)\ ,\\
	\zb &\,=\, ie^{-i\tau(u)}\left(1-\epsilon\sigma(u)+\cO(\epsilon^2)\right)\ . \label{eq: bosonic wiggling coordinate}
\end{aligned}
\end{equation}
Demanding that the induced metric on the wiggling boundary is 
\begin{equation}
	\frac{du^2}{\epsilon^2}\,=\,4e^{2\Sigma}\z'(u)\zb'(u) du^2 \ ,
\end{equation}
we have 
\begin{equation}
	\sigma(u)\,=\,\tau'(u)\ .
\end{equation}
Furthermore, one can obtain the tangential vector and normal vector of the wiggling boundary:
\begin{equation}
	\tangen^\z\,=\,\z'(u)\ ,\qquad \normal_\z\,=\,{i\over 1-\z(u)\zb(u)}\sqrt{\zb'(u)\over\z'(u)}\ ,
\end{equation}
where we normalize the normal vector by $\normal^m\normal_m=1$. From the the normal vector and the tangential vector, the extrinsic curvature is found to be
\begin{equation}
	K\,=\,1+\epsilon^2\left(\mathrm{Sch}\left[\tau;u\right]+{\tau'(u)^2\over 2}\right) + \cO(\epsilon^4) \ ,
\end{equation}
where the Schwarzian derivative $\mathrm{Sch}(\tau;u)$ is defined as
\begin{equation}
	\mathrm{Sch}\left[\tau;u\right]\,=\,  {\tau'''(u)\over \tau'(u)}-{3\over 2}\left({\tau''(u)\over \tau'(u)}\right)^2\ .
\end{equation}
Furthermore, the dilaton solution on the wiggling boundary is 
\begin{align}
    \phi\,=\, {1 +z\bar{z} \over 1- z \bar{z}}\,=\, {1\over \epsilon\tau'}+\cO(\epsilon^0) \ .
\end{align}
Then the action of JT gravity boils down to the boundary action
\begin{equation}
	I_{\text{JT}}\,=\,2\int\displaylimits_{\partial\mathcal{M}}\frac{\dd u}{\tau'(u)} \,\left(\mathrm{Sch}\left[\tau;u\right]+{\tau'(u)^2\over 2}-\frac{1}{2}\right)\ , 
\end{equation}
where we used $\sqrt{h}={1\over\epsilon}+\cO(\epsilon^0)$ and the counterterm $K_0=\left.K\right\vert_{\tau(u)=u}=1+{\epsilon^2\over2}$. Then using the inversion formula of Schwarzian derivative
\begin{equation}
    \mathrm{Sch}\left[\tau;u\right]\,=\,-\tau'(u)^2\mathrm{Sch}\left[u;\tau\right]\ ,
\end{equation}
we finally obtain the Schwarzian action at finite temperature ($\beta=2\pi$)~\cite{Joung:2023doq}
\begin{equation}
	I_{\text{JT}}\,=\,-2\int\displaylimits_{\partial\mathcal{M}}\dd \tau \,\left(\mathrm{Sch}\left[u;\tau\right]+{u'(\tau)^2\over 2}-\frac{1}{2}\right)\ . \label{eq: bosonic Schwarzian action after inversion}
\end{equation}
By the inversion formula, the overall sign of the Euclidean action is flipped. In fact, the final Euclidean action \eqref{eq: bosonic Schwarzian action after inversion} with overall minus sign is stable~\cite{Witten:1987ty}. Furthermore, the inversion from $\tau(u)$ into $u(\tau)$ gives the path integral measure of the Schwarzian action in~\cite{Bagrets:2016cdf,Stanford:2017thb,Cotler:2018zff,Joung:2023doq}
\begin{equation}
    \cD \tau(u) \,=\, \frac{\cD u(\tau)}{u'(\tau)}\ .
\end{equation}

\subsection{Frame Field Fluctuation}
\label{sec:vielbein}
%%%%%%%%%%%%%%%%%%%%%%%%%%%%%%%%%%%%%%%%%%%%%%%%%%%%%%%%%%%

The gravitational edge mode can also be described by the metric fluctuation. In $\ads_2$, one can find a coordinate transformation which gives the exact $\ads_2$ metric while it deforms the full $\ads_2$ boundary. Under such a coordinate transformation, the metric fluctuation is absorbed to the wiggling boundary~\cite{Joung:2023doq}.

In the metric-like formulation, the radial diffeomorphism is broken on the boundary, and therefore one has to restrict the radial diffeomorphism on the boundary. Accordingly, the degrees of freedom, which would have been gauged away by the boundary radial diffeomorphism, becomes physical when the broken radial diffeomorphism is disallowed. In this section we will study this would-be gauge mode, or in other words the gravitational edge mode, in the frame-like formulation where the frame field fluctuation on the boundary leads to the edge mode as in the metric-like formulation~\cite{Joung:2023doq}. In particular we will consider a coordinate transformation preserving the conformal gauge which induces the broken radial diffeomorphism. And this coordinate transformation corresponds to the gravitational edge mode.

Let us first clarify our conventions for frame-like formulation. The flat and curved coordinates are denoted by bold and normal fonts respectively. For example $\bw\ , \bwb$ denote flat coordinates, and $w \ , \wb$ denote curved coordinates. We also collectively denote curved and flat coordinates as $x^{m}$ (lower case middle Latin letters) and $x^{a}$ (lower case beginning Latin letters) respectively.

The conformal gauge for the frame field $e\indices{_{m}^{a}}$ is defined by
\begin{equation}
e\indices{_{w}^{\bw}} \,=\, e\indices{_{\wb}^{\bwb}}  \ , \qquad e\indices{_{w}^{\bwb}} \,=\, e\indices{_{\wb}^{\bw}} \,=\, 0 \ . \label{eq: conformal gauge in bosonic jt}
\end{equation}
From the background solution~\eqref{eq:BosBG} of the bulk equation of motion, the solution for the frame field $e\indices{_{m}^{a}}$ in conformal gauge is found to be\footnote{The flat metric is normalized with a factor of $4$ to match the supergravity conventions in~\cite{Chamseddine:1991fg,Forste:2017kwy}.} 
\begin{equation}
		\dd s^{2} \,=\, 4\,  \dd w \dd \wb\, \, e\indices{_w^{\bw}}  e\indices{_{\wb}^{\bwb}}\;\;,\quad   e^{2 \Omega} \,=\,  e\indices{_w^{\bw}} \,=\, e\indices{_{\wb}^{\bwb}} \,=\,   \frac{1}{1 - w \wb } \ .\label{eq: frame solution bosonic jt}
\end{equation}
To introduce the would-be gauge modes, we consider a coordinate transformation between the background coordinates $(w,\wb)$ and new coordinates $(z,\zb)$: 
\begin{equation}
z \;\;\longrightarrow \;\; w \,=\, W(\z,\zb)   \;\; , \qquad  \zb \;\;\longrightarrow \;\; \wb \,=\, \overline{W}(\z,\zb)\ .
\end{equation}
This coordinate transformation induces the transformation of the frame fields:
\begin{equation}\label{eq:BosFrPB}
e\indices{_{m}^{a}} \;\;\longrightarrow \;\; \tilde{e}\indices{_{n}^{b}} \,\equiv\, J\indices{_{n}^{m}}e\indices{_{m}^{a}} \;  , \quad J\indices{_{n}^{m}} \,\equiv\, \pdv{w^{m}}{z^{n}} = \frac{\del (w ,\wb) }{\del (z,\zb)}   \ ,
\end{equation}
where $w^m=(w,\wb)$ and $z^n=(z,\zb)$. Since we work within the conformal gauge~\eqref{eq: conformal gauge in bosonic jt}, we demand that the transformed frame fields satisfy
\begin{equation}\label{eq:BosGauge}
\tilde{e}\indices{_{z}^{\bz}} \,=\, \tilde{e}\indices{_{\zb}^{\bzb}} \ , \qquad 	\tilde{e}\indices{_{z}^{\bzb}} \,=\, \tilde{e}\indices{_{\zb}^{\bz}} \,=\, 0 \ .
\end{equation}
The diagonality of frame fields imposed by conformal gauge \eqref{eq:BosGauge} imply the holomorphicity of the coordinate transformation
\begin{alignat}{2}
	0&\,=\,\tilde{e}\indices{_{\z}^{\bzb}}\,=\,\partial_\z \wb\, e\indices{_{\wb}^{\bwb}}&&\quad\Longrightarrow\quad \partial_\z \wb\,=\,0\ ,\\
	0&\,=\,\tilde{e}\indices{_{\zb}^{\bz}}\,=\,\partial_{\zb} \w\, e\indices{_{\w}^{\bw}}&&\quad\Longrightarrow\quad \partial_{\zb} \w\,=\,0\ .
\end{alignat}
As in the wiggling boundary point of view, to get a Schwarzian description, we restrict to small fluctuations near the boundary of the $\ads_2$ background. Such holomorphic functions can be asymptotically expanded near the boundary in terms of polar coordinates. Near the boundary, with $r = \sqrt{\z\zb} = 1-\epsilon$ and $\epsilon\ll1$, we have
\begin{align}
\begin{split}\label{eq:BosDiffeoExp}
\z \to \w & \,\equiv\, W(\z)\,=\,-i e^{i \tau(u)}\bigg(1-\epsilon \tau'(u)+\frac{\epsilon^2}{2}\big( \tau'(u)^2- \tau'(u)-i \tau''(u)\big)+\cO\left(\epsilon^3\right)\bigg) \ , 
\end{split}\\
\begin{split}
    \zb \to \wb & \,\equiv\, \overline{W}(\zb)\,=\,i e^{-i \tau(u)}\bigg(1-\epsilon \tau'(u)+\frac{\epsilon^2}{2}\big( \tau'(u)^2- \tau'(u)+i \tau''(u)\big)+\cO\left(\epsilon^3\right)\bigg) \ , 
\end{split}
\end{align}
where the leading terms correspond to the boundary diffeomorphism $\tau(u)$. We also demand the monodromy condition 
\begin{align}
    \tau(u+2\pi)\,=\, \tau(u)+2\pi \ 
\end{align}
on $\tau$  to ensure that the geometry does not have a conical singularity at $r = 0$. 

We find that the above transformation fails to preserve the conformal gauge as the transformed frame violates the reality condition, \textit{i.e.}, we have $\tilde{e}\indices{_{z}^{\bz}} \ne \tilde{e}\indices{_{\zb}^{\bzb}}$. However, note that in addition to diffeomorphisms, the frame-like formulation also enjoys the local Lorentz symmetry parameterized by real boost angle $\Lambda$.  Its action is given by
\begin{equation}
\tilde{e}\indices{_{m}^{a}} \;\;\longrightarrow \;\; \tilde{e}\indices{_{m}^{a}}L\indices{_{a}^{b}} \ , \quad L\indices{_{a}^{b}} \,\equiv\, \text{diag}(e^{i \Lambda}, e^{-i\Lambda} ) \ .
\end{equation}
Therefore, to preserve the conformal gauge, on top of the diffeomorphism \eqref{eq:BosDiffeoExp}, we also invoke a compensating local Lorentz transformation. More explicitly, under the action of  $L\indices{_{a}^{b}}$, the diagonal components of frame field transform as
\begin{equation}
    \tilde{e}\indices{_{\z}^{\bz}}\,=\,\partial_\z W(z)\, e\indices{_{\w}^{\bw}}e^{i\Lambda}\ , \qquad \tilde{e}\indices{_{\zb}^{\bzb}}\,=\,\partial_{\zb} \overline{W}(\zb)\, e\indices{_{\wb}^{\bwb}}e^{-i\Lambda}\ ,
\end{equation}
where the appropriate boost angle $\Lambda$ is determined to be
\begin{equation}
	\Lambda\,=\,u-\tau(u)-\epsilon{\tau''(u)\over \tau'(u)}+{\epsilon^2\over2}\tau''(u)(1-{\tau'(u)}^{-1})+\cO(\epsilon^3)\ .
\end{equation}
Finally, we get the real-valued frame field:
\begin{equation}
\begin{aligned}
	\tilde{e}\indices{_{z}^{\bz}} \,=\, \tilde{e}\indices{_{\zb}^{\bzb}} &\,=\,{1\over2\epsilon}+{1\over4}-{\epsilon\over6}\left( \mathrm{Sch}\left[\tau;u\right]+{\tau'(u)^2\over 2} -{5\over4}\right)+\cO(\epsilon^2) \ .
\end{aligned}
\end{equation}
From the frame field fluctuation, we can obtain the metric fluctuation
\begin{equation}
	\dd s^{2} \,=\, 4\, \tilde{e}\indices{_{z}^{\bz}} \tilde{e}\indices{_{\zb}^{\bzb}}\, \dd\z\dd\zb\,=\, 4e^{2\widetilde{\Omega}} (\dd r^{2}+r^{2} \dd u^{2}) \ ,
\end{equation}
where the conformal factor $\widetilde{\Omega}$ is
\begin{equation}
		e^{2 \widetilde{\Omega}}  \,=\,  {1\over4\epsilon^2}+{1\over4\epsilon}-{1\over6}\left( \mathrm{Sch}\left[\tau;u\right]+{\tau'(u)^2\over 2} -{13\over8}\right)+\cO(\epsilon)\ .
\end{equation}
Moreover, the extrinsic curvature on the boundary $r = 1$ is written as
\begin{equation}
K \,=\, 1 + \epsilon^{2} \left( \mathrm{Sch}\left[\tau;u\right]+{\tau'(u)^2\over 2} \right)+ \cO(\epsilon^4)\ .
\end{equation}
As we have done in the previous section, one can derive the Schwarzian action that describes the dynamics of edge mode on the boundary after inverting the variable $\tau(u)$ into $u(\tau)$:
\begin{equation}
	I_{\text{JT}}\,=\,-2\int\displaylimits_{\partial\mathcal{M}}\dd \tau \,\left(\mathrm{Sch}\left[u;\tau\right]+{u'(\tau)^2\over 2}-\frac{1}{2}\right)\ .
\end{equation}
This boundary action from the frame field fluctuation is identical to that from wiggling boundary prescription in Eq.~\eqref{eq: bosonic Schwarzian action after inversion} .

%%%%%%%%%%%%%%%%%%%%%%%%%%%%%%%%%%%%%%%%%%%%%%%%%%%%%%%%%%%
%%%%%%%%%%%%%%%%%%%%%%%%%%%%%%%%%%%%%%%%%%%%%%%%%%%%%%%%%%%
\section{\texorpdfstring{$\cN=1$}{N=1} Jackiw-Teitelboim Supergravity and Edge Mode}
\label{sec: super jt gravity}
%%%%%%%%%%%%%%%%%%%%%%%%%%%%%%%%%%%%%%%%%%%%%%%%%%%%%%%%%%%
%%%%%%%%%%%%%%%%%%%%%%%%%%%%%%%%%%%%%%%%%%%%%%%%%%%%%%%%%%%

The $\cN=1$ supersymmetric Jackiw-Teitelboim (sJT) gravity action including the boundary action term can be written as \cite{Forste:2017kwy} 
\begin{equation}
\begin{aligned}
    I_{sJT} &\,=\, i \int_{\cM} d\w d\wb\, d\etp d\etm\, \sfr\, \Phi\left(\cR-2\right)+ 2\int_{\partial \cM} du d \bth\, \mathsf{e}\,\Phi (\cK-\cK_0)\ , \label{eq: sJT}
\end{aligned}
\end{equation}
where the bulk two-dimensional bosonic space $(\w,\wb)$ is extended to the superspace $(\w,\wb,\etp,\etm)$. Hence we use superframe $\sfr\indices{_{M}^{A}}$ to write in the first-order formalism with its Berezinian $\sfr$, and $\mathsf{e}$ denotes the Berezinian of boundary superframe. Moreover, dilaton and the geometric quantities such as Riemann curvature and extrinsic curvature are promoted to the supersymmetric one: dilaton superfield $\Phi$, super-curvature $\cR$ and ``super-extrinsic curvature'' $\cK$. See Appendix~\ref{app: Conventions for supergravity} for the details about the conventions and definitions of supergeometric quantities.

The line element on the super geometry can be expressed in terms of the superframe field $\sfr\indices{_{M}^{A}}$ as
\begin{equation}
\dd s^{2} \,=\, 4 \left(\dd x^{M} \sfr\indices{_{M}^{\bw}} \right) \left(\dd x^{N}\sfr\indices{_{N}^{\bwb}}\right) \ . \label{eq: super line element}
\end{equation}
In order to proceed, we first fix the gauge redundancies in $\sfr\indices{_M^{A}}$. Note that the $(2|2)$-dimensional supergeometry is completely characterized by its superconformal factor $e^{\swe}$~\cite{Martinec:1983um,Chamseddine:1991fg}, which should in turn determine all components of the superframe. Following \cite{Martinec:1983um}, we  work with the superconformal gauge where all quantities of interest split into holomorphic and anti-holomorphic terms, with the latter given by complex conjugation of the former. In what follows, we  explicitly report only the holomorphic expressions. The superconformal gauge is given by
\begin{equation} \label{eq:sucongauge}
	\begin{split}
			\sfr_{\bep} \,=\, e^{-\frac{1}{2} \swe}\Dp \ , \quad \sfr_{\bw} \,=\, e^{\swe}\left\{e^{-\frac{1}{2}\swe}\sfr_{\bep} , e^{-\frac{1}{2}\swe}\sfr_{\bep} \right\} \; , \qquad (\Dp \equiv \dep+\etp \dw)\ .
	\end{split}
\end{equation}
where $(\bw,\bwb,\bep,\bem)$ denote the coordinates for the tangent space. We describe the evaluation of the superframe more explicitly in Appendix~\ref{app: Conventions for supergravity}. To summarize, the holomorphic block of the superframe $\sfr\indices{_{A}^{M}}$ is given by 
\begin{equation}\label{eq:ConfGaugeSuFrame1}
	\sfr\indices{_{A}^{M}}|_{\text{holo}} \,=\,	\left[\begin{array}{c|c}
			\sfr\indices{_{\bw}^{w}} & \sfr\indices{_{\bw}^{\etp}} \\
			\hline
			\sfr\indices{_{_\bep}^{w}} & \sfr\indices{_{_\bep}^{\etp}} 
		\end{array} \right] \, =\, 
		 \left[ \begin{array}{c|c}
			\Dp\big( \etp e^{-\swe}\big) & \Dp\big( e^{-\swe} \big)	 \\
			\hline 
			\etp e^{-\frac{1}{2}\swe}& e^{-\frac{1}{2}\swe}
		\end{array}\right]  \ , 
\end{equation}  
and its inverse
\begin{equation}\label{eq:ConfGaugeSuFrame2}
	\sfr\indices{_{M}^{A}}|_{\text{holo}}	\,=\,	 \left[ \begin{array}{c|c}
		\sfr\indices{_{w}^{\bw}} & \sfr\indices{_{w}^{\bep}} \\
		\hline
		\sfr\indices{_{\etp}^{\bw}} & \sfr\indices{_{\etp}^{\bep}} 
	\end{array} \right] \,=\,  \left[ \begin{array}{c|c}
		e^{\swe} &  e^{-\frac{1}{2} \swe} \Dp e^{ \swe}   \\
		\hline
		- \etp e^{\swe} &  e^{-\frac{1}{2} \swe} \Dp\big(\etp e^{ \swe} \big)
	\end{array}\right] \ .
\end{equation}
The full superframe takes the block diagonal form $\sfr\indices{_{M}^{A}} = \text{diag}(\sfr\indices{_{M}^{A}}|_{\text{holo}},\sfr\indices{_{M}^{A}}|_{\text{anti-holo}})$. Hence, the line element \eqref{eq: super line element} at superconformal gauge \eqref{eq:sucongauge} is given by
\begin{equation}
	\dd s^{2} \,=\, 4 e^{2\swe} \left( \dd w + \etp \dd \etp \right)\left( \dd \wb + \etm \dd \etm \right) \  .  \label{eq: super line element at superconformal gauge}
\end{equation}
In the conformal gauge, the associated super-curvature $R_{\bep \bem}$\footnote{This is the super-curvature component denoted as $R_{_{+-}}$ in \cite{Forste:2017kwy,Chamseddine:1991fg}.} can be expressed as (See \cite{Forste:2017kwy,Chamseddine:1991fg}):
\begin{equation}
	\cR \equiv R_{\bep\bem} = -2 i e^{-\swe}\Dp\Dm\swe \ . \label{eq: supercurvature}
\end{equation}
As in the case of the bosonic JT gravity, the dilaton superfield plays a role of Lagrange multiplier to impose the bulk constraint $\cR=2$. Using Eq.~\eqref{eq: supercurvature}, this constraint becomes
\begin{align}
    \Dp\Dm\swe \,=\, i e^{\swe}\ .
\end{align}
This equation can be solved by\footnote{Our convention for complex conjugation of Grassmann numbers \textit{does not} change the order of product. With this convention, the superconformal factor is real.}
\begin{equation}
	e^{\swe} \,=\, \frac{1}{1-w \wb + i \etp \etm}  \ . \label{eq: super poincare conformal factor}
\end{equation}
Note that we have selected out the solution for $\swe$ which corresponds to the super-Poincare disk geometry. This conformal factor reduces to that of the Poincare disk metric \eqref{eq:BosBG} after truncating the Grassmann odd coordinates $\etp,\etm$.

In superconformal gauge with super-Poincere disc geometry, the variation with respect to the superframe gives the equations of motion for the dilaton superfield:
\begin{equation}
	\Dp\Dm\Phi \,=\, ie^{\swe}\Phi\ , \quad \Dp\left[e^{-\swe}\Dp\left(e^{-\swe}\Dp\Phi\right)\right]\,=\,0\ ,\quad  \Dm\left[e^{-\swe}\Dm\left(e^{-\swe}\Dm\Phi\right)\right]\,=\,0 \ ,
\end{equation}
and the dilaton superfield solution can be found to be
\begin{equation}
    \Phi\,=\,\frac{1+\w\wb}{1-\w\wb+i\etp\etm}\ .
\end{equation}

The background metric defined with the superconformal factor \eqref{eq: super poincare conformal factor} has the $PSU(1,1|1)$ isometry given by
\begin{equation}
			\begin{split}
				w &\to \frac{p w  - i  \nu\etp + q}{\bar{q}w+\bar{\nu}\etp + \bar{p}} \ , \qquad 	\etp \to \frac{ \mu w +e \etp + i \bar{\mu}}{\bar{q}w+\bar{\nu}\etp + \bar{p}}  \ , \\ \wb &\to \frac{\bar{p} \wb  + i  \bar{\nu}\etm + \bar{q}}{q\wb+{\nu}\etm + p} \ , \qquad 	\etm \to \frac{ \bar{\mu} \wb +e \etm - i {\mu}}{q \wb+{\nu}\etm + p} \ ,
			\end{split} \label{eq: bulk psu isometry}
\end{equation}
where
\begin{equation}
        \begin{aligned}
			&\mu \,=\, \bar{\nu} p  + i \nu \bar{q} \ , \quad \bar{\mu} \,=\, {\nu} \bar{p}  - i \bar{\nu} q \ ,\\
            &e - i \nu \bar{\nu} \,=\,1 \ , \quad |p|^{2} - |q|^{2}+i\nu \bar{\nu} \,=\, 1 \ .
        \end{aligned}
\end{equation}
The $SU(1,1|1)$ group structure can be seen manifestly by a group element defined as
\begin{equation}
\left[\begin{array}{c c |c}
\bar{p} & \bar{q} & \bar{\nu}\\
q & p & - i\nu \\
\hline
i \bar{\mu} & \mu & e
\end{array} \right]\ ,
\end{equation}
where it satisfies 
\begin{equation}
\left[\begin{array}{c c |c}
p & \bar{q} & - i\mu\\
q & \bar{p} & \bar{\mu}\\
\hline
-\nu & -i \bar{\nu} & e
\end{array} \right]\left[\begin{array}{c c |c}
-1 & 0 & 0\\
0 & 1 & 0\\
\hline
0 & 0 & i
\end{array} \right] \left[\begin{array}{c c |c}
\bar{p} & \bar{q} & \bar{\nu}\\
q & p & - i\nu \\
\hline
i \bar{\mu} & \mu & e
\end{array} \right] = \left[\begin{array}{c c |c}
-1 & 0 & 0\\
0 & 1 & 0\\
\hline
0 & 0 & i
\end{array} \right] \ .
\end{equation}
Furthermore, under the $PSU(1,1|1)$ isometry the dilaton $\Phi$ transforms as
\begin{equation}
\Phi \rightarrow \frac{(p w  - i  \nu\etp + q)(\bar{p} \wb  + i  \bar{\nu}\etm + \bar{q})+(\bar{q}w+\bar{\nu}\etp + \bar{p})(q\wb+{\nu}\etm + p)}{1 - \w \wb +i \etp \etm} \ .
\end{equation}

%%%%%%%%%%%%%%%%%%%%%%%%%%%%%%%%%%%%%%%%%%%%%%%%%%%%%%%%%%%
%%%%%%%%%%%%%%%%%%%%%%%%%%%%%%%%%%%%%%%%%%%%%%%%%%%%%%%%%%%
\subsection{Wiggling Boundary}
\label{sec:supersymmetric wiggling boundary}
%%%%%%%%%%%%%%%%%%%%%%%%%%%%%%%%%%%%%%%%%%%%%%%%%%%%%%%%%%%
%%%%%%%%%%%%%%%%%%%%%%%%%%%%%%%%%%%%%%%%%%%%%%%%%%%%%%%%%%%

In this section, we extend our prescription for the derivation of the Schwarzian action with wiggling boundary in Section~\ref{sec: wiggling boundary in conformal gauge} to get the super-Schwarzian action at finite temperature. We begin with the exact super-Poincare disk geometry as the base manifold with the metric
\begin{equation}
    ds^2\,=\, 4e^{2\Sigma}\vert d\w+\etp d\etp\vert^2\ ,\qquad e^\Sigma\,=\,\frac{1}{1-\w\wb+i\etp\etm}\ .
\end{equation}
Thus, we intrinsically begin with the the finite temperature setting and the derivation at zero temperature can be found in Ref.~\cite{Forste:2017kwy}. 

We begin with a parametrization of the boundary wiggling close to the full super-Poincare disk ($r\equiv\sqrt{\w\wb}\equiv 1-\epsilon$) by\footnote{Factor $i$ multiplied in $z,\bar z$ is our convention to simply write Eq.~\eqref{eq: eq1 from Dz=etaDeta} without additional factors.}
\begin{equation}
\begin{aligned}
	\z&\,=\,-ie^{i\tau(u)}(1-\epsilon\sigma+\cO(\epsilon^2))\ ,\\
	\zb &\,=\,ie^{-i\tau(u)}(1-\epsilon\sigma+\cO(\epsilon^2))\ ,\\
	\tp&\,=\,e^{\frac{i\tau(u)}{2}}(\xi+\epsilon\rho+\cO(\epsilon^2))\ ,\\
	\tm&\,=\,e^{-\frac{i\tau(u)}{2}}(\xi+\epsilon\bar\rho+\cO(\epsilon^2))\ , \label{eq: wiggling coordinate}
\end{aligned}
\end{equation}
where $\sigma(u)$ and $\rho(u)$ are real Grassmann even and complex Grassmann odd functions, respectively. $\sigma$ and $\rho$ can be solved in terms of $\xi(u)$ and its derivatives by demanding the boundary superframe to be
\begin{equation}
\begin{aligned}
	\frac{(du+\bth d\bth)^2}{\epsilon^2}&\,=\,4e^{2\Sigma}\left\vert\frac{\partial \z}{\partial u}+\tp\frac{\partial\tp}{\partial u}\right\vert^2(du+\bth d\bth)^2\\
 &\quad-4e^{2\Sigma}\bigg[\left(\frac{\partial\zb}{\partial u}+\tm\frac{\partial\tm}{\partial u}\right)(\bD z-\tp \bD\tp)+\left(\frac{\partial\z}{\partial u}+\tp\frac{\partial\tp}{\partial u}\right)(\bD \zb-\tm \bD\tm) \bigg]d\bth du\ ,\label{eq: boundary metric identification}
\end{aligned}
\end{equation}
where the boundary super-derivative is defined by $\bD \equiv \del_{\bth} + \bth \del_{u}$.

Now we solve the map to wiggling boundary \eqref{eq: wiggling coordinate} up to order $\cO(\epsilon^1)$ and determine $\sigma$ and $\rho$ therein. Using the expansion of the superconformal factor near the boundary
\begin{equation}
	e^\Sigma=\frac{1}{2\epsilon\sigma}\left(1-\frac{i\xi(\bar\rho-\rho)}{2\sigma}\right) +\cO(\epsilon^0)\ ,
\end{equation}
one can solve Eq.~\eqref{eq: boundary metric identification} perturbatively:%\footnote{To get the form of Eq.~\eqref{eq: du component matching}, the relation $\partial_u \z+\tp\partial_u\tp=(\bD\tp)^2$ is utilized, where Eq.~\eqref{eq: component proportionality condition} is used.}
\begin{align}
	&\bD\z\,=\,\tp \bD\tp\ ,\quad \bD\zb\,=\,\tm \bD\tm\ ,\label{eq: component proportionality condition}\\
	&\frac{1}{4\epsilon^2}\,=\,e^{2\Sigma} \left(\bD\tp \bD\tm\right)^2\ .\label{eq: du component matching}
\end{align}
And this leads to an equation for $\sigma$:
\begin{align}
	\sigma\,=\,\left(1+\frac{i\xi(\rho-\bar\rho)}{2\sigma}\right)(\bD\xi)^2+\cO(\epsilon)\ .
\end{align}
The perturbative solution is found to be
\begin{equation}
	\sigma\,=\,(\bD\xi)^2+{i\xi(\rho-\bar{\rho})\over2}+\cO(\epsilon)\ .\label{eq: sigma sol1}
\end{equation}
Moreover, Eq.~\eqref{eq: component proportionality condition} can be expressed perturbatively in terms of $\sigma, \rho$ and $\xi$ as follows.
\begin{align}
	\bD\tau&=\xi \bD\xi\ , \label{eq: eq1 from Dz=etaDeta}\\
	-\xi \bD\xi\sigma+i\bD\sigma&=\xi \bD\rho+\rho \bD\xi\ .\label{eq: eq2 from Dz=etaDeta}
\end{align}
Using Eq.~\eqref{eq: sigma sol1}$\sim$\eqref{eq: eq2 from Dz=etaDeta}, we find 
%
%\begin{equation}
%	-\xi (\bD\xi)^3+2i\bD\xi \bD^2\xi-i\bD\xi\rho_2+i\xi \bD\rho_2\,=\,\xi \bD\rho_1+\rho_1 \bD\xi+i\xi \bD\rho_2+i\rho_2 \bD\xi\ .
%\end{equation}
%
%Separating the real and imaginary parts of the equations gives the solution of $\rho\equiv\rho_1+i\rho_2$ and in turn, it finally determines $\sigma$ in terms of $\xi$ as
%
\begin{equation}
    \sigma\,=\,(\bD\xi)^2-\xi \bD^2\xi\ ,\qquad \rho\,=\,-\frac12\xi(\bD\xi)^2+i\bD^2\xi\ .
\end{equation}

Armed with the above relations, we may now evaluate the boundary action on the wiggling boundary. The ``super-extrinsic curvature''\footnote{This definition of the ``super-extrinsic curvature'' was proposed by~\cite{Forste:2017kwy,Forste:2017apw}. It remains an open question whether this ``super-extrinsic curvature'' leads to the well-defined variational principle like Gibbons-Hawking term.} $\mathcal{K}$ is defined by
\begin{equation}
\cK \equiv \frac{\tangen^{\bb{a}}\bD_{\tangen} \normal_{\bb{a}}}{\tangen^{\bb{b}}\tangen_{\bb{b}}}  \ , \qquad \bD_{\tangen}\equiv \bD z^{M} \bDp_{M}\big|_{\text{bdy}}  \ , \quad  \tangen^{\bb{a}}\equiv \del_{u} z^{M}\sfr\indices{_{M}^{\bb{a}}}\big|_{\text{bdy}} \ . \label{eq: superextrinsic curvature}
\end{equation}
Here $\tangen^{\bb{a}}$ denotes the tangent vector to the boundary (more precisely its projection along the bosonic flat directions) and $\normal_{\bb{a}}$ the normal vector to the boundary. $\bDp_{M}$ denotes the covariant derivative described in Appendix~\ref{app: Conventions for supergravity}.

To compute the super-extrinsic curvature~\eqref{eq: superextrinsic curvature}, we first obtain the tangential vector $\tangen^A$ and the normal vector $\normal_A$ of the wiggling boundary.
\begin{equation}
	\tangen^\bz\,=\, \frac{1}{2\epsilon}\frac{\bD\tp}{\bD\tm}\; ,\quad \normal_\bz\,=\,\frac{i\bD\tm}{\bD\tp}\ ,
\end{equation}
where the normal vector is normalized by $\normal_\bz \normal_{\bzb}=1$. Then the super-extrinsic curvature~\eqref{eq: superextrinsic curvature} and the corresponding counterterm $\cK_0$ can be evaluated to be 
\begin{equation}
\cK\,=\,2\epsilon^2\left(\ssch[\tau,\xi;u,\bth]+\frac14\xi(\bD\xi)^3\right)+\cO(\epsilon^3)\; ,\quad \cK_0\,\equiv\,\cK\,\big|_{\tau(u)=u\,,\,\xi(u)=\bth}\,=\,{\epsilon^2\bth \over 2} . \label{eq: extrinsic curvature with wiggling}
\end{equation}
where super-Schwarzian (sSch) \cite{Fu:2016vas} is defined by
\begin{equation}
	\ssch[\tau,\xi;u,\bth]\,=\,\frac{\bD^4\xi}{\bD\xi}-2\frac{\bD^2\xi \bD^3\xi}{(\bD\xi)^2}\ .\label{eq: super-Schwarzian}
\end{equation}
Note that the super-extrinsic curvature $\cK$ does not have a term of order $\cO(\epsilon^0)$, which would have given rise to the divergence on the boundary in contrast to the bosonic JT gravity.

On the boundary, the dilaton superfield solution becomes
\begin{equation}
\Phi \quad\longrightarrow\quad \frac{1}{\epsilon (\bD\xi)^2}\ ,\label{eq: boundary super-dilaton}
\end{equation}
and in Ref.~\cite{Forste:2017kwy} the boundary super-volume was argued to be of the form\footnote{It is still unclear to us how to obtain two components of the boundary superframe in order to confirm this argument. These are not necessary in evaluating the extrinsic curvature, and therefore it does not have any influence on the rest of calculations. }
\begin{equation}
	\int du d\bth\,\mathsf{e} \quad\longrightarrow\quad \frac{1}{\epsilon}\int du d\bth\ . \label{eq: boundary measure}
\end{equation}
Therefore, we find
\begin{equation}
\begin{aligned}
	I_{\text{sJT}} &\,=\, 4\int du d\bth\, (\bD\xi)^{-2}\, \bigg(\ssch[\tau,\xi;u,\bth]+\frac{\xi}{4}(\bD_\bth\xi)^3-\frac14\bth\bigg) +\cO(\epsilon)\ . \label{eq: super schwarzian before inversion}
\end{aligned}
\end{equation}
Due to the factor $(\bD\xi)^{-2}$ which comes from the dilaton superfield, the resulting action is not seemingly the super-Schwarzian action. As in Section~\ref{sec: wiggling boundary in conformal gauge}, we need supersymmetric version of the inversion formula~\cite{Fu:2016vas,Cardenas:2018krd}:
\begin{align}
	\mathrm{Sch}[\tau,\xi;u,\bth]&\,=\,-(\bD\xi)^{3}\mathrm{Sch}[u,\bth;\tau,\xi]\ ,\\
    du d\bth&\,=\,d\tau d\xi\, (\bD\xi)^{-1}\ .
\end{align}
After inverting $\xi(u,\bth)$ into $\bth(\tau,\xi)$, we have
\begin{equation}
\begin{aligned}
	I_{\text{sJT}} &\,=\, -4\int d\tau d\xi\,  \bigg(\ssch[u,\bth;\tau,\xi]+\frac{\bth}{4}(\bD_{\xi}\bth)^3-\frac14\xi\bigg) +\cO(\epsilon)\ , \label{eq: super schwarzian after inversion}
\end{aligned}
\end{equation}
where $\bD_\xi\equiv \partial_\xi+\xi\partial_\tau$ and $(\bD\xi)^{-1}=\bD_\xi\bth$. Note that the second term $\frac{\bth}{4}(\bD_{\xi}\bth)^3$ is the finite temperature contribution to the super-Schwarzian action, and it comes from the counterterm $\cK_0$. On the other hands, the term $\frac14\xi(\bD\xi)^3$ in Eq.~\eqref{eq: super schwarzian before inversion} becomes $-\frac14\bth$ after the inversion.

%%%%%%%%%%%%%%%%%%%%%%%%%%%%%%%%%%%%%%%%%%%%%%%%%%%%%%%%%%%
%%%%%%%%%%%%%%%%%%%%%%%%%%%%%%%%%%%%%%%%%%%%%%%%%%%%%%%%%%%
\subsection{Superframe Field Fluctuation}
\label{sec: superframe field fluctuation}
%%%%%%%%%%%%%%%%%%%%%%%%%%%%%%%%%%%%%%%%%%%%%%%%%%%%%%%%%%%
%%%%%%%%%%%%%%%%%%%%%%%%%%%%%%%%%%%%%%%%%%%%%%%%%%%%%%%%%%%

Now we can further extend the description of the gravitational edge mode by frame field fluctuation in Section \ref{sec:vielbein} to the supersymmetric one. We start by introducing another super-Poincare disk geometry with coordinates $\z, \zb, \tp, \tm$ and corresponding flat coordinates $\bz,\bzb, \btp, \btm$. They are collectively denoted as $\z^{N}$ and $\z^{B}$ respectively. The corresponding $(2|2)$-dimensional super-derivatives are written as
\begin{equation}
\Dpn \equiv \dtp + \tp \dz \ , \quad \Dmn \equiv \dtm + \tm \dzb \ .
\end{equation}
As in the bosonic case, we consider a coordinate transformation between $(w,\etp,\wb,\etm)$ for the background $\ads_2$ and $(\z,\tp,\zb,\tm)$ for the other super-Poincare disk:
\begin{equation}
\begin{split}
\z \;\;\longrightarrow\;\; \w &\,=\, W(\z,\tp,\zb,\tm)   \ , \qquad   \tp \;\;\longrightarrow\;\; \etp\,=\,\Theta(\z,\tp,\zb,\tm)\ ,  \\
\zb \;\;\longrightarrow\;\; \wb &\,=\, \overline{W}(z,\tp,\zb,\tm)   \ , \qquad   \tm \;\;\longrightarrow\;\; \etm\,=\,\overline{\Theta}(z,\tp,\zb,\tm)  \ .\\
\end{split}
\end{equation}
The pullback of the superframe field is then given by
\begin{equation}
	\begin{split}
	&\sfr\indices{_{M}^{A}}  \quad\longrightarrow\quad   \nsfr\indices{_{N}^{B}} \,=\, \mathsf{J}\indices{_{N}^{M}} \sfr\indices{_{M}^{A}} \mathsf{L}\indices{_{A}^{B}}  \ ,  \\
	&\mbox{where}\quad \mathsf{J}\indices{_{N}^{M}}  \,\equiv\, \frac{\del \w^{M}}{\del \z^{N} } = \frac{\del (\w,\etp,\wb,\etm)}{\del(\z,\tp,\zb,\tm)} \ , \qquad \mathsf{L}\indices{_{A}^{B}}  \,\equiv\, \text{diag}(e^{i\srap}, e^{\frac{i}{2} \srap}, e^{-i\srap}, e^{-\frac{i}{2} \srap})  \ .
	\end{split}
\end{equation}
together with the local Lorentz gauge transformation $\mathsf{L}$. This local Lorentz transformation will ensure that the superframe fields are real.
Now we need to demand that the diffeomorphism $W, \overline{W},\Theta,\overline{\Theta}$ preserves the superconformal gauge~\eqref{eq:sucongauge}. This requires the transformed superframe $ \nsfr\indices{_{N}^{B}}$ to be block diagonal, which leads to the holomorphic mappings, \textit{viz}.,
\begin{equation}
\w = W(\z,\tp), \quad \etp = \Theta(\z,\tp), \quad \wb = \overline{W}(\zb, \tm), \quad \etm = \overline{\Theta}(\zb,\tm) .
\end{equation}
To preserve the superconformal gauge, we also require to satisfy the relation
\begin{equation}
	\tp \nsfr\indices{_{z}^{\bz}} +  \nsfr\indices{_{\tp}^{\bz}} = 0  \ .
\end{equation}
Demanding the above imposes an effective $(1|1)$-dimensional super diffeomorphism condition on the holomorphic coordinates via
\begin{equation}\label{eq:HolSuDiff}
	\Dpn W(\z,\tp) = \Theta(\z,\tp) \Dpn \Theta(\z,\tp)\ , 
\end{equation}
and likewise for the maps $\overline{W}$ and $\overline{\Theta}$. Thus we arrive at the form
\begin{equation}
\begin{aligned}
	\left[\begin{array}{c|c}
	\nsfr\indices{_{z}^{\bz}} & \nsfr\indices{_{z}^{\btp}} \\
	\hline
	\nsfr\indices{_{\tp}^{\bz}} & \nsfr\indices{_{\tp}^{\btp}}
\end{array} \right] &\,=\, \left[ \begin{array}{c|c}
\left( \Dpn \Theta \right)^{2} e^{\swe} &e^{-\frac{1}{2} \swe}  \Dpn\big(\Dpn \Theta \,  e^{\swe} \big) \\
- \tp \left( \Dpn \Theta \right)^{2} e^{\swe} & e^{-\frac{1}{2} \swe}\Dpn\big(\tp \, \Dpn \Theta \,  e^{\swe} \big) \\
\end{array} \right]\left[\begin{array}{c c}
e^{i\srap } & 0 \\
0 & e^{\frac{i}{2}\srap} \
\end{array} \right] \ ,
\end{aligned}
\end{equation}
and similarly for the conjugate anti-holomorphic block.\footnote{To simplify expressions we make use of the identities following from the relation \eqref{eq:HolSuDiff}:
\begin{equation}
\dz W + \Theta \dz \Theta \,=\,  \left(\Dpn \Theta \right)^{2} \ , \quad \dtp W - \Theta \dtp \Theta \,=\, - \tp \left( \Dpn \Theta\right)^{2} \ .
\end{equation} } 
Finally we need to impose the reality condition. This is easily achieved by fixing the compensating Lorentz rotation to be
\begin{equation}
e^{i\srap} \,=\, \frac{\Dmn \overline{\Theta}}{\Dpn \Theta} \ . \label{eq: super-Lorentz transformation}
\end{equation}
Similarly, the transformed super spin-connection is obtained as
\begin{equation}
	\begin{split}
		\widetilde{\spc}_{N} \,=\, \mathsf{J}\indices{_{N}^{M}} \spc_{M} - \del_{N} \srap , 
	\end{split}
\end{equation}
giving explicitly,
\begin{equation}
\widetilde{\spc}_{\z} \,=\, - i \dz \left(\swe + \log \Dpn \Theta \right) \ , \  \widetilde{\spc}_{\tp} \,=\, - i \dtp\left( \swe + \log \Dpn \Theta \right) \ , \label{eq: transformed super spin-connection form}
\end{equation}
and for the anti-holomorphic components via complex conjugation. Finally, notice that the induced super-Poincare disk geometry is characterized by the superconformal factor
\begin{equation}
	e^{\widetilde{\swe}} \,=\, e^{\swe} \left(\Dpn \Theta \Dmn \overline{\Theta}\right) \,  \ ,
\end{equation}
in the superconformal gauge. The above results are well-known in the literature on $(2|2)$-dimensional super geometry \cite{Chamseddine:1991fg, Martinec:1983um} and we review them here for completeness. The upshot is that the super diffeomorphisms which preserve the superconformal gauge are fully characterized as explained above.

%%
%Our purpose is to find the residual physical degrees of freedom to be appeared as edge mode that conserve the gauge condition \eqref{eq: block superframe gauge}. For the bosonic case, allowed transformations are holomorphic transformation $\z\rightarrow f(\z)$ and antiholomorphic transformation $\zb\rightarrow\bar f(\zb)$ which are two-dimensional conformal transformations. 

Holomorphic transformations $W,\Theta$ satisfying Eq.~\eqref{eq:HolSuDiff} are found to be finite superconformal transformations \cite{Chamseddine:1991fg,Arvis:1982tq} 
\begin{equation}
\begin{aligned}
	W(\z,\tp)&\,=\,F\left(\z+\tp\Psi(\z)\right)\,=\,F(\z)+\tp\Psi(\z)\dz F(\z)\\
	\Theta(\z,\tp)&\,=\,\sqrt{\dz F(\z)}\left(\tp+\Psi(z)+\frac{\tp}{2}\Psi(\z)\dz\Psi(\z)\right)\ , \label{eq: 2D superspace diffeomorphism}
\end{aligned}
\end{equation}
where $F(\z)$ and $\Psi(\z)$ are functions with even and odd Grassmann parities respectively. Near the boundary, these functions can be determined as a radial expansion after adopting a polar decomposition $\z = -i r e^{i u}$ and $\zb = i r e^{-i u}$. Furthermore, we define the boundary polar coordinates by
\begin{equation}
    \left.\z\right\vert_{\text{bdy}}\,=\,-ie^{iu}\ ,\quad  \left.\tp\right\vert_{\text{bdy}}\,=\,e^{iu\over2}\bth\ , \quad \left.\zb\right\vert_{\text{bdy}}\,=\,ie^{-iu}\ ,\quad  \left.\tm\right\vert_{\text{bdy}}\,=\,e^{-iu\over2}\bth\ .
\end{equation}
Using $\tau(u)$ and $\psi(u)$ which are respectively bosonic and fermionic functions on the boundary, we have
\begin{equation} 
\begin{aligned}
    F&\,=\,-i e^{i \tau(u)}\bigg(1-\epsilon \tau'(u)+\frac{\epsilon^2}{2}\left( \tau'(u)^2- \tau'(u)-i \tau''(u)\right)+\cO\left(\epsilon^3\right)\bigg)\ ,\\
    \Psi&\,=\, e^{\frac{i\tau(u)}{2}}\bigg(\psi(u)+\epsilon\left(i\psi'(u)-\frac{\psi(u)}{2}\right)-\frac{\epsilon^2}{2}\left(\psi''(u)+\frac{\psi(u)}{4}\right) +\cO\left(\epsilon^3\right)\bigg)\ .\label{eq: 2D superspace diffeomorphism form}
\end{aligned}
\end{equation}
where $r=\sqrt{\z\zb}\equiv 1-\epsilon$. The leading terms $(\epsilon\rightarrow0)$ in Eq.~\eqref{eq: 2D superspace diffeomorphism form} are designed to write the transformed coordinate on the boundary in polar-decomposed form
\begin{equation}
    \left.\w\right\vert_{\text{bdy}}\,=\,-ie^{i\upsilon}\ ,\quad  \left.\etp\right\vert_{\text{bdy}}\,=\,e^{i\upsilon\over2}\xi\ , \quad \left.\wb\right\vert_{\text{bdy}}\,=\,ie^{-i\upsilon}\ ,\quad  \left.\etm\right\vert_{\text{bdy}}\,=\,e^{-i\upsilon\over2}\xi\ ,
\end{equation}
with boundary bosonic coordinate $\upsilon$ and boundary fermionic coordinate $\xi$. Then, from Eq.~\eqref{eq: 2D superspace diffeomorphism}, we have
\begin{equation}
    \upsilon\,=\,\tau\left(u+\bth\psi(u)\right)%\,=\,\tau(u)+\bth\psi(u)\tau'(u)
    \; ,\;\; \xi\,=\,\sqrt{ \tau'(u)}\left(\bth+\psi(u)+\frac12\bth\psi(u)\psi'(u)\right)\ . \label{eq: boundary superfield}
\end{equation}
This turns out to be the solution of the one-dimensional super-diffeomorphism condition~\cite{Fu:2016vas}, $\bD\upsilon=\xi\bD\xi$ in Eq.~\eqref{eq: eq1 from Dz=etaDeta}.

Equipped with the above results, we now proceed to calculate the super-extrinsic curvature. The tangential vector and unit normal vector on the boundary of the super-Poincare disk are 
\begin{equation}
	\tangen^\bz\,=\,\frac{e^{iu}}{2\epsilon}+\cO(\epsilon^0)\ ,\qquad \normal_\bz\,=\,ie^{-iu}+\cO(\epsilon)\ .
\end{equation}
Evaluating the super-extrinsic curvature~\eqref{eq: superextrinsic curvature} with the above gives:
\begin{equation}
\begin{aligned}
	\cK\,=\,&\epsilon^{2}\bigg[\left(2+\psi\psi'\right)\psi''+\psi\left(\mathrm{Sch}[\tau;u]+\frac12\tau'^2\right)\\
	&+\bth\bigg(-1+\psi\psi'''+3\psi'\psi''+(1-\psi\psi')\left(\mathrm{Sch}[\tau;u]+\frac12\tau'^2\right)\bigg)\bigg]+\cO(\epsilon^3)\ ,
\end{aligned}
\end{equation} 
and the counterterm is given by
\begin{equation}
    \cK_0\,\equiv\,\cK\,\big|_{\tau(u)=u\,,\,\xi(u)=\bth}\,=\,-{\epsilon^2\bth \over 2}\ .
\end{equation}
In terms of superfield $\xi$ in Eq.~\eqref{eq: boundary superfield}, we have
\begin{equation}
\begin{aligned}
\cK-\cK_0&\,=\,2\epsilon^2\left(\ssch[\tau,\xi;u,\bth]+\frac14\xi(\bD\xi)^3-\frac14\bth\right)+\cO(\epsilon^3)\ .
\end{aligned}
\end{equation}
By the same variable inversion formula $\xi(u,\bth)\rightarrow\bth(\tau,\xi)$ used in Section~\ref{sec:supersymmetric wiggling boundary}, we obtain the same boundary action for the $\cN=1$ supersymmetric gravitational edge mode \eqref{eq: super schwarzian after inversion} 
\begin{equation}
	I_{\text{sJT}} \,=\, -4\int d\tau d\xi\,  \bigg(\ssch[u,\bth;\tau,\xi]+\frac{\bth}{4}(\bD_{\xi}\bth)^3-\frac14\xi\bigg) +\cO(\epsilon)\ .\label{eq: schwarzian action superframe fluctuation}
\end{equation}
One can easily check that the truncation of fermionic variables reduce this boundary action into the bosonic Schwarzian action \eqref{eq: bosonic Schwarzian action after inversion}.

%%%%%%%%%%%%%%%%%%%%%%%%%%%%%%%%%%%%%%%%%%%%%%%%%%%%%%%%%%%
\subsection{\texorpdfstring{$OSp(2|1)$}{OSp(2|1)} Gauging in Super-Schwarzian Theory}
\label{sec: osp gauging}
%%%%%%%%%%%%%%%%%%%%%%%%%%%%%%%%%%%%%%%%%%%%%%%%%%%%%%%%%%%

The super-Schwarzian action in Eq.~\eqref{eq: schwarzian action superframe fluctuation} is invariant under the $OSp(2|1)$ transformation~\cite{Fu:2016vas}. However, this $OSp(2|1)$ transformation should not be regraded as a symmetry of the boundary action. Rather, it is a gauge symmetry which eliminates zero modes in the super-Schwarzian theory. It was pointed out in Ref.~\cite{Maldacena:2016upp,Alkalaev:2022qfc,Choi:2023syx,Joung:2023doq} that this gauge symmetry originates from the isometry of the background $AdS_2$ geometry. 

The origin of the gravitational edge mode is the broken radial diffeomorphism, and therefore we can parametrize this edge mode by a large diffeomorphism with respect to a reference geometry, for example, the background $AdS_2$ geometry. Hence the isometry of the background $AdS_2$ geometry causes the redundancy in the parameterization of the edge mode.

\begin{figure}[t!]
\centering
\begin{tikzpicture}
\draw[domain=0:2*pi,samples=200,smooth, variable=\x,fill=black!5] plot ({(2.0)*cos(deg(\x))},{(2.0))*sin(deg(\x))});

\draw[domain=0:2*pi,samples=200,smooth, variable=\x] plot ({6.8+(2.0)*cos(deg(\x))},{(2.0))*sin(deg(\x))});
\draw[domain=0:2*pi,samples=200,smooth, variable=\x,fill=black!5] plot ({6.8+(1.7+0.02*(1*cos(deg(3*\x))+2*cos(deg(7*\x))+3*cos(deg(19*\x))+4*cos(deg(11*\x))))*cos(deg(\x))},{(1.8+0.01*(1*cos(deg(3*\x))+2*cos(deg(7*\x))+3*cos(deg(19*\x))+4*cos(deg(11*\x))))*sin(deg(\x))});

\node[circle,fill=black,scale=0.4, label=below:{$(\w,\etp,\wb,\etm)$}] (n1) at (0.8,0.6) {};

\node[circle,fill=black,scale=0.4, label=below:{$(\z,\tp,\zb,\tm)$}] (n2) at (7.5,0.8) {};

\draw[-latex] (n1) to[bend left]node[midway, sloped, above]{immersion} (n2);
%\draw[-latex,bend left]  (n1) edge (n2);
\node[] at (0,-2.5) {Base super-Poincare Disk};
\node[] at (6.8,-2.5) {Target super-Poincare Disk};

\end{tikzpicture}
	\caption{The mapping (\textit{immersion}) from the base super-Poincare disk to the target super-Poincare disk can be interpreted as the gravitational edge mode. }
	\label{fig: immersion}
\end{figure}
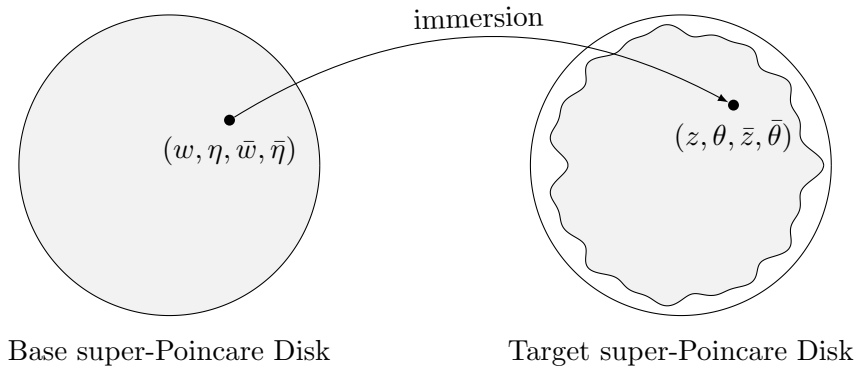 
The inversion procedure in the derivation of the super-Schwarzian plays an crucial role in elucidating the relation between the isometry and the $OSp(2|1)$ gauging~\cite{Joung:2023doq}. To single out the edge mode in Section~\ref{sec: super jt gravity}, we have employed a map -- or more precisely \textit{immersion}~\cite{Ferrari:2024kpz} -- between \textit{base super-Poincare disk} $(\w,\etp,\wb,\etm)$ and \textit{target super-Poincare disk} $(\z,\tp,\zb,\tm)$ (See Figure~\ref{fig: immersion}). The immersion from base super-Poincare disk to the target super-Poincare disk parameterizes the edge mode, and the isometry of the base super-Poincare disk leads to the redundant description of the edge mode\footnote{Equivalently, one can introduces a Stueckelberg field to address the gravitational edge mode, thereby recovering the full super-diffeomorphism on the boundary\cite{Joung:2023doq}. Then, after fixing the superframe to be the exact AdS$_2$, the residual gauge symmetry, which corresponds to the isometry of AdS$_2$, also leads to the $OSp(2|1)$ gauging.}~\cite{Joung:2023doq,Ferrari:2024kpz}. Consequently, it is necessary to apply the corresponding equivalence relation to the boundary coordinates $(u,\bth)$ of the base super-Poincare disk. And this equivalence relation among the boundary coordinates $(u,\bth)$ becomes the $OSp(2|1)$ gauging in the super-Schwarzian theory via inversion procedure.

We emphasize that the isometry of the base super-Poincare does not necessarily have the same form as the $OSp(2|1)$ gauging although the isometry generates the $OSp(2|1)$ gauging. To understand this distinction, let us first consider the boundary isometry of the base super-Poincare disk. Using the expansion of the coordinate transformation in Eq.~\eqref{eq: 2D superspace diffeomorphism form}, one can obtain the boundary isometry acting on the boundary coordinate $(u,\bth)$ from the bulk isometry group of the exact super-Poincare disk geometry~\eqref{eq: bulk psu isometry} becomes :
\begin{align}
\tan \frac{u}{2}\quad&\longrightarrow\quad\frac{a \tan \frac{u}{2}+b-\gamma \frac{1}{\sqrt{2}} \sec \frac{u}{2} \bth}{c \tan \frac{u}{2}+d-\alpha \frac{1}{\sqrt{2} }\sec \frac{u}{2} \bth}\ , \label{eq: bosonic super isometry}\\
 -\frac{1}{\sqrt{2}} \sec \frac{u}{2} \bth\quad&\longrightarrow\quad\frac{\beta+\delta \tan \frac{u}{2}-e \frac{1}{\sqrt{2}} \sec \frac{u}{2} \bth}{c \tan \frac{u}{2}+d-\alpha \frac{1}{\sqrt{2}} \sec \frac{u}{2} \bth}\ . \label{eq: fermionic super isometry}
\end{align}
where the $PSU(1,1|1)$ parameters is mapped to the $OSp(2|1)$ parameters as follows.
\begin{equation}
\begin{split}
p & =\frac{1}{2}(a+d+i(b-c))\ ,\qquad  q=\frac{1}{2}(b+c+i(a-d))\ , \\
\nu & = - \frac{1}{\sqrt{2}}\left(\alpha + i\gamma \right) \ ,\qquad \qquad \mu =  - \frac{1}{\sqrt{2}}\left(\delta + i\beta \right) \ ,
\end{split}
\end{equation}
with the conditions
\begin{equation}
\begin{split}
ad-bc+\alpha \gamma &=1\ , \quad e-\alpha \gamma =1 \ , \\
a \alpha - c \gamma &=\delta \ , \quad b\alpha -d \gamma = \beta \ .  
\end{split}
\end{equation}

On the other hand, in the super-Schwarzian action~\eqref{eq: schwarzian action superframe fluctuation}, we use the one-dimensional superdiffeomorphism condition for the boundary coordinates $(u,\bth)$:
\begin{equation}
    u\,=\,  U\big(\upsilon + \xi\chi(\upsilon)\big)\;\;, \quad \bth \,=\,  \sqrt{\partial_\upsilon  U(\upsilon)} \bigg( \xi + \chi(\upsilon) + {1\over 2} \xi \chi(\upsilon) \chi'(\upsilon)  \bigg)\ . \label{eq: one dim superdiffeo condition 2}
\end{equation}
We find that this one-dimensional super-diffeomorphism condition can be obtained from the same condition for $(\upsilon, \xi)$ in Eq.~\eqref{eq: boundary superfield} by 
\begin{align}
    \chi(\upsilon)\,=\,& - {1\over \sqrt{\partial_\upsilon  U(\upsilon)}} \psi \big( U(\upsilon)\big)\ ,\\
    U(\upsilon)\,= \,& \tau^{-1}(\upsilon)\ .
\end{align}
Therefore, the super-Schwarzian action in Eq.~\eqref{eq: schwarzian action superframe fluctuation} is a functional of $U$ and $\chi$ rather than of $u$ and $\bth$. Thus, the isometry acts on the boundary coordinates $(u,\bth)$ while the $OSp(2|1)$ gauging acts on the function $(U,\chi)$. In addition, the $OSp(2|1)$ gauging is induced by the isometry in Eq.~\eqref{eq: bosonic super isometry} and Eq.~\eqref{eq: fermionic super isometry} via Eq.~\eqref{eq: one dim superdiffeo condition 2}:
\begin{align}
	\tan {U\over 2} \, \sim\,&       {a \tan {U\over 2} +b  + \gamma \sqrt{{U' \over 2}} \sec {U \over 2} \chi \over c\tan {U\over 2} +d + \alpha \sqrt{{U' \over 2}}  \sec{U \over 2} \chi  } \ ,\label{eq: osp gauge jt 1}\\
	\sqrt{{U' \over 2}} \sec{U \over 2}\chi \,\sim \, &      { \delta \tan {U\over 2} + \beta + e \sqrt{{U' \over 2}} \sec{u \over 2} \chi   \over  c \tan {U\over 2} + d+ \alpha \sqrt{{U' \over 2}} \sec{U \over 2} \chi }\ .\label{eq: osp gauge jt 2}
\end{align}
Note that this $OSp(2|1)$ transformation for $(U,\chi)$ is distinct from the isometry in Eq.~\eqref{eq: bosonic super isometry} and Eq.~\eqref{eq: fermionic super isometry}, primarily because the former involves $U'$ while the latter does not. Furthermore, from the $OSp(2|1)$ gauging in the zero-temperature Schwarzian action, one can derive the $OSp(2|1)$ gauging at finite temperature via a transformation $f(\tau)=\tan{\tau\over 2}$~\cite{Fu:2016vas}. The one-dimensional super-diffeomorphism condition for the transformation $f(\tau)=\tan{\tau\over 2}$ also produces $U'$ factors, leading to results that are identical to those in Eq.~\eqref{eq: osp gauge jt 1} and Eq.~\eqref{eq: osp gauge jt 2}.

%%%%%%%%%%%%%%%%%%%%%%%%%%%%%%%%%%%%%%%%%%%%%%%%%%%%%%%%%%%
%%%%%%%%%%%%%%%%%%%%%%%%%%%%%%%%%%%%%%%%%%%%%%%%%%%%%%%%%%%
\section{\texorpdfstring{$osp(2|1)$}{osp(2|1)} BF Description}
\label{sec: bf description}
%%%%%%%%%%%%%%%%%%%%%%%%%%%%%%%%%%%%%%%%%%%%%%%%%%%%%%%%%%%
%%%%%%%%%%%%%%%%%%%%%%%%%%%%%%%%%%%%%%%%%%%%%%%%%%%%%%%%%%%

The $\cN=1$ JT supergravity can be formulated by the $osp(2|1)$ BF theory~\cite{Fan:2021wsb} of which the action is given by
\begin{equation}
	S_{BF} \,=\, \int_{\mathcal{M}} \str \big( \mathbf{\Phi} \mathbf{F} \big)  \  ,\label{eq: bulk action bf}
\end{equation}
where $\mathbf{\Phi}$ is a 0-form field and $\mathbf{F} \equiv d \mathbf{A} + \mathbf{A}\wedge\mathbf{A} $ is a 2-form field strength. In terms of $osp(2|1)$ generators, $L_n$ and $G_\alpha$ ($n=1,0,-1$ and $\alpha=\pm{1\over 2}$), (see Appendix~\ref{app: osp algebra} for details), they can be written as
\begin{equation}
	\mathbf{\Phi} \,\equiv \, \left[\begin{array}{c c|c}
    {1\over 2}\Phi_0 & - \Phi_{-1} & \Phi_{-{1\over 2}}\\
    \Phi_1 & -{1\over 2}\Phi_0 & \Phi_{{1\over 2}} \\
    \hline
    -\Phi_{{1\over 2}} & \Phi_{-{1\over 2}} & 0\\
    \end{array}\right]  \ , \quad \mathbf{A} \,\equiv\, \left[\begin{array}{c c|c}
    {1\over 2}A_0 & - A_{-1} & A_{-{1\over 2}}\\
    A_1 & -{1\over 2}A_0 & A_{{1\over 2}} \\
    \hline
    -A_{{1\over 2}} & A_{-{1\over 2}} & 0\\
    \end{array}\right] \ .
\end{equation}

For the Euclidean supergravity of $\ads_2$, we consider the BF action on the disk $\cM$. Hence we need to add the boundary action $S_{bdy}$ on the boundary of the disk $\cM$ for the well-defined variational principle:
\begin{align}
    S_{bdy}\,\equiv\, \kappa\int_{\partial \cM} d \tau \; \str \bA_\tau^2 \ , \label{eq: boundary action bf}
\end{align}
where $\tau$ denotes the coordinate along the boundary $\partial\cM$. This boundary term breaks the gauge symmetry on the boundary, which leads to the gravitational edge mode in the BF theory~\cite{Joung:2023doq}. To avoid the broken gauge symmetry on the boundary, we may restrict the gauge parameter on the boundary. This gives rise to physical degrees of freedom on the boundary which would have been gauged away by an unrestricted gauge parameter. This physical degrees of freedom corresponds to the gravitational edge mode or would-be gauge mode~\cite{Balachandran:1994up,Balachandran:1995qa,Arcioni:2002vv,Carlip:1995cd,Takayanagi:2019tvn,Donnelly:2020teo,David:2022jfd,Mertens:2022ujr,Wong:2022eiu,Joung:2023doq,Mukherjee:2023ihb}.

Equivalently, the edge mode can be introduced by the (unrestricted) gauge transformation~\cite{Joung:2023doq}. Since the boundary action is not invariant under the unrestricted gauge transformation, the gauge parameter remains in the boundary action. And we can promote the gauge parameter on the boundary to the dynamical edge mode like Stuekelberg field, thereby clearly separating the edge mode from the bulk field. In this formulation the variation with respect to the edge mode yields the equation of motion of the edge mode while the variation with respect to the bulk field leads to the boundary condition of the bulk field.

In this paper, we will consider the restricted gauge transformation on the boundary for the edge mode, which is seemingly simpler than the separation of bulk field and edge mode. However one need careful treatment on gauge field $\mathbf{A}$ which includes the physical edge mode on the boundary. For example, the variation of the total action $S_{tot}\equiv S_{BF} + S_{bdy}$ is 
\begin{align}
    \delta S_{tot} \,=\, \text{(E.o.m)} + \int_{\partial \mathcal{M}} \str\big[ \big( \mathbf{\Phi}+ \kappa \mathbf{A}\big)\delta \mathbf{A}\big] \ .
\end{align}
It was demonstrated in Ref.~\cite{Joung:2023doq} that the boundary condition $\mathbf{\Phi}+ \kappa \mathbf{A}|_{\partial \cM}=0$ or $\delta \mathbf{A}|_{\partial \cM}=0$ on the boundary kills the edge mode. This is because the variation with respect to the edge mode, which is the equation of motion of the edge mode, was taken as a boundary condition in the path integral. Hence, to see the well-defined variational principle, one had better divide the gauge field $\mathbf{A}$ into the bulk field $\mathbf{A}_{\text{\tiny bulk}}$ and the edge mode $k_{\text{\tiny edge}}$, (\eg $\mathbf{A}\,=\, k_{\text{\tiny edge}}^{-1}\mathbf{A}_{\text{\tiny bulk}}k_{\text{\tiny edge}}+k_{\text{\tiny edge}}^{-1}d k_{\text{\tiny edge}}$). We can then take the Dirichlet boundary condition for the bulk field, $\delta \mathbf{A}_{\text{\tiny bulk}}|_{\partial \cM}=0$ for the well-defined variational principle.

We choose the gauge condition analogous to the Fefferham-Graham gauge in the metric-like formulation:
\begin{align}
    \mathbf{A}_r\,=\, b^{-1}(r)\partial_r b(r) \qquad \mbox{where}\quad b(r)\,\equiv \, e^{r L_0}\ .
\end{align}
This gauge condition is valid because one can find a gauge transformation, which is trivial on the boundary, and changes arbitrary $\mathbf{A}_r$ to the form $b^{-1}(r)\partial_r b(r)$~\cite{Campoleoni:2010zq}. With this gauge condition, one may write $\mathbf{A}_\tau$ as
\begin{align}
    \mathbf{A}_\tau \,=\, b^{-1}\, \mathbf{a}(\tau)\, b \ .
\end{align}
Note that the flat connection condition, which is the bulk constraint imposed by the bulk dilaton field, implies that $\mathbf{a}$ is a function of $\tau$ only.

The (boundary) transformation of $\mathbf{a}$
\begin{align}
    \mathbf{a}\quad\longrightarrow \quad  h^{-1}\, \mathbf{a}\, h +  h^{-1}\,\partial_u h \ .\label{eq: boundary transformation of a}
\end{align}
changes the boundary action if the gauge parameter $h(\tau)$ is non-trivial, and therefore it is not an allowed gauge transformation in this formulation. Hence, the boundary gauge fields $\mathbf{a}$'s which are connected by the non-trivial gauge parameter $h$ are the physically non-redundant degrees of freedom. These physical configurations can be counted by one reference configuration $\mathbf{a}_\ads$ and the \textit{smooth}\footnote{Here, ``smooth'' means that the gauge parameter $h$ is smoothly connected to the identity matrix.} gauge parameter $h$:
\begin{align}
    \mathbf{a}\,=\, h^{-1}\,\mathbf{a}_\ads\, h+ h^{-1}\partial_\tau h \ .
\end{align}
This gauge parameter $h$ corresponds to the gravitational edge mode or would-be gauge mode. By using the transformation\footnote{Or, if we consider $\mathbf{a}_\ads$ and $h$ as dynamical fields, we may recover the boundary gauge transformation, $\mathbf{a}_\ads\to k^{-1} \mathbf{a}_\ads k +k^{-1}\partial_\tau k $ and $h\to k^{-1} h$ . Using this gauge transformation by $k$, one may fix $\mathbf{a}_\ads$. }~\eqref{eq: boundary transformation of a}, the reference configuration $\mathbf{a}_\ads$ can be chosen to be
\begin{align}
    \mathbf{a}_\ads\,=\, \left[\begin{array}{c c|c}
    0 & - \mathcal{L}_0 & 0\\
    1 & 0 & 0 \\
    \hline
   0 & 0 & 0\\
    \end{array}\right]
\end{align}
where $\cL_0$ is a constant which is related to the Casimir of $\mathbf{a}$, and therefore one cannot eliminate it by the smooth transformation.

In the $sl(2,\mathbb{R})$ BF formulation for the bosonic JT gravity, the asymptotic AdS condition can be incorporated by the first class constraint imposed by the nilpotent subgroup~\cite{Valach:2019jzv,Joung:2023doq}. Together with the gauge condition for the first class constraint, one has
\begin{align}
    \mathbf{a}\,=\,\begin{pmatrix}
        {1\over 2} a_0 & -\mathcal{L}(\tau)\\
        a_1 & -{1\over 2}a_0 \\
    \end{pmatrix}\,=\,\begin{pmatrix}
        0 & -\mathcal{L}(\tau)\\
        1 & 0 \\
    \end{pmatrix}
\end{align}
where $a_1=1$ and $a_0=0$ is the first class constraint and the gauge condition, respectively.

For the $osp(2|1)$ BF theory, the asymptotic AdS condition in the JT supergravity can be translated into~\cite{Henneaux:1999ib,Cardenas:2018krd,Fan:2021wsb}
\begin{align}
    \mathbf{a}\,=\,\left[\begin{array}{c c|c}
    {1\over 2}a_0 & - a_{-1} & \psi_{-{1\over 2}}\\
    a_1 & -{1\over 2}a_0 & \psi_{{1\over 2}} \\
    \hline
    -\psi_{{1\over 2}} & \psi_{-{1\over 2}} & 0\\
    \end{array}\right]\,=\,  \left[\begin{array}{c c|c}
    0 & - \mathcal{L}(\tau)  & \mathcal{G}(\tau)\\
    1 & 0 & 0 \\
    \hline
   0 &\mathcal{G}(\tau) & 0\\
    \end{array}\right]\ .\label{eq: asymptotic ads condition in bf}
\end{align}
However it is not straightforward to impose this condition at the level of Lagrangian as a constraint. For example, if the $\psi_{1\over 2}=0$ is the first class constraint, we do not have the corresponding gauge condition in Eq.~\eqref{eq: asymptotic ads condition in bf}. In addition, one can also observe the difficulty by comparing with the Hamiltonian reduction in the constrained WZW model where it is natural to impose the following constraints on the WZW currents $J^a$~\cite{Bershadsky:1989mf}.
\begin{align}
    J^{-a}\,=\, \begin{cases}
    \quad 1 \qquad & \mbox{for}\;\;a\;:\;\mbox{positive simple root} \\
    \quad 0 \qquad & \mbox{for}\;\;a\;:\;\mbox{positive non-simple root}\\
    \end{cases}
\end{align}
In the asymptotic AdS condition~\eqref{eq: asymptotic ads condition in bf} for $osp(2|1)$ BF formulation, $a_1$ is not related to the simple root of $osp(2|1)$ while $\psi_{1\over2}$ corresponds to the simple root. Hence one cannot follow the same Hamiltonian reduction as in the $SL(2,\mathbb{R})$ BF formulation.

For the case of $osp(2|1)$ case, it was shown in Ref.~\cite{Bershadsky:1989tc} that one need to introduce extra fermion field which will be eliminated as a gauge condition of the first class constraint $\psi_{1\over 2}=0$. We revisit this procedure in Appendix~\ref{app: constraint in bf theory}, but we still do not have a framework to impose those constraints at the Lagrangian level.

To understand the global structure, it is useful to define a Wilson line from $\tau=0$ to $\tau=2\pi$:
\begin{align}
    \cW[\mathbf{a}]\,\equiv\,\cP \exp\bigg( \int_0^{2\pi} \mathbf{a} \;d\tau  \bigg)\ .
\end{align}
Here, we choose the ordering of the matrix exponential such that the value of $\tau$ is increasing  from left to right of the product. Writing the boundary gauge field $\mathbf{a}$ in Maurer-Cartan form
\begin{align}
\mathbf{a}\,=\, g^{-1} \partial_\tau g \qquad \mbox{where} \;\; g\in OSp(2|1) \ , \label{eq: Maurer-Cartan form}
\end{align}
the Wilson line becomes
\begin{align}
    \cW[\mathbf{a}]\,=\, g^{-1}(0) g(2\pi) \ . 
\end{align}
Hence the monodromy relation of $g(\tau)$ reads
\begin{align}
    g(2\pi)\,=\, g(0) \cW(\mathbf{a})\ .
\end{align}
The monodromy matrix $\cW(\mathbf{a})$ was classified by the conjugacy class of the $OSp(2|1)$ group in Ref.~\cite{Fan:2021wsb}. We are interested in the special Elliptic II class where the holonomy is given by
\begin{align}
    \cW(\mathbf{a})\,=\, \left[\begin{array}{c c | c}
  -1 & 0 & 0 \\
  0 & -1 & 0\\
  \hline
  0 & 0 & 1
\end{array} \right] \ .
\end{align}
Note that the monodromy matrix of the background gauge field $\mathbf{a}_\ads$ is 
\begin{align}
    \cW(\mathbf{a}_\ads)
    \,= \, %\exp \left(2\pi\left[\begin{array}{c c|c}
%    0 & - \mathcal{L}_0 & 0\\
%    1 & 0 & 0 \\
%    \hline
%   0 & 0 & 0\\
%    \end{array}\right]\right)\,= \, 
    \, \left[\begin{array}{c c|c}
    \cos (2\pi \sqrt{\cL_0}) & -\sqrt{\cL_0}\sin (2\pi \sqrt{\cL_0}) & 0\\
    {1\over \sqrt{\cL_0}}\sin (2\pi \sqrt{\cL_0}) & \cos (2\pi \sqrt{\cL_0}) & 0 \\
    \hline
   0 & 0 & 1\\
    \end{array}\right] \ ,
\end{align}
which determines $\cL_0\,=\, {1\over 4}$. For this class of the monodromy matrix, one can use the \textit{Iwasawa-like} decomposition of the $OSp(2|1)$ group element $g$: 
\begin{align}
		g\, \equiv \,&  \left[\begin{array}{c c | c}
			\cos \frac{ u(\tau)}{2} & -\sin \frac{ u(\tau)}{2}  & 0\\
			\sin \frac{ u(\tau)}{2} & \cos \frac{ u(\tau)}{2} & 0 \\
			\hline
			0 & 0 & 1 \\
		\end{array}\right]  \left[ \begin{array}{c c|c}
	 [y(\tau)]^{-{1\over 2}} & 0 & 0\\
	 0 & [y(\tau)]^{1\over 2} & 0\\
	 \hline
	 0 & 0 & 1
     \end{array}\right] \cr
     &\times\left[\begin{array}{c c | c}
  	                      1 & 0 & 0 \\
  	                      0 & 1 & - \bth(\tau)\\
  	                      \hline
  	                      \bth(\tau) & 0 & 1
  \end{array} \right]\left[\begin{array}{c c | c}
  1 & 0 & \eta(\tau) \\
  0 & 1 & 0\\
  \hline
  0 & \eta(\tau) & 1
\end{array} \right] \left[ \begin{array}{c c|c}
	 1 & - f(\tau) & 0\\
	 0 & 1 & 0\\
	 \hline
	 0 & 0 & 1
     \end{array}\right]\ ,\label{eq: iwasawa-like decomposition}
\end{align}
where $u(\tau)$, $f(\tau)$, $y(\tau)$ are bosonic coordinates while $\bth(\tau)$ and $\eta(\tau)$ denote fermionic ones. Furthermore in this paper we will consider the NS sector in which $u(\tau)$ is a function of $\tau$ with winding number 1; $y(\tau)$ and $f(\tau)$ is a periodic function; $\bth(\tau)$ and $\eta(\tau)$ is an anti-periodic function. \ie
\begin{align}
    u(\tau+2\pi)\,=\,& u(\tau)+ 2\pi\ ,\cr
    y(\tau+2\pi)\,=\,& y(\tau)\;,\quad f(\tau+2\pi)\,=\, f(\tau)\ ,\cr
    \bth(\tau+2\pi)\,=\,& -\bth(\tau)\;\quad \eta(\tau+2\pi )\,=\, -\eta(\tau) \ .
\end{align}
One can easily confirm that the monodromy relation\footnote{Note that the monodromy relation is different from that in Ref.~\cite{Fan:2021wsb}. The Iwasawa-like decomposition in this work can capture the monodromy structure of the $osp(2|1)$ BF theory on the disk, which naturally leads to the finite temperature super-Schwarzian. On the other hand, the solution found in Ref.~~\cite{Fan:2021wsb} is involved with the zero-temperature super-Schwarzian, and therefore the (large gauge) transformation to the finite temperature will modify the monodromy relation.} for the Iwasawa-like decomposition~\eqref{eq: iwasawa-like decomposition} is found to be
\begin{align}
    g(2\pi) \,=\, -  g(0)(-)^F \ ,
\end{align}
where the matrix $(-)^F$ is defined by
\begin{align}
    (-)^F\,\equiv\, \left[\begin{array}{c c | c}
  1 & 0 & 0 \\
  0 & 1 & 0\\
  \hline
  0 & 0 & -1
\end{array} \right] \ .
\end{align}
Using the Iwasawa decomposition~\eqref{eq: iwasawa-like decomposition}, one can solve the asymptotic condition~\eqref{eq: asymptotic ads condition in bf}. We find
\begin{align}
    y\,=\,&  \frac{1}{2} u'\big(1+ \bth  \bth' \big)\ ,\\
    f\,=\,&   \frac{1}{2} \bigg({u''\over u'}+\bth \bth''\bigg)\ ,\\
    \eta\,=\,& \bth' + \frac{1}{2}\bth  {u''\over u'}\ ,
\end{align}
and 
\begin{align}
    \mathcal{L}\,=\,& \frac{1}{2} \big(1- \bth  \bth' \big)\bigg(\sch(u,\tau) +\frac{1}{2} u'^2\bigg) + \frac{1}{2}\big(\bth \bth''' + 3\bth' \bth''\big) \ ,\\
    \mathcal{G}\,=\,& \frac{1}{2} \bth \bigg(\sch(u,\tau) +\frac{1}{2} u'^2 \bigg) + \bth'' + \frac{1}{2}\bth  \bth'  \bth''\ .
\end{align}
With this solution, the bulk action $S_{BF}$ vanishes due to the constraint while $S_{bdy}$ gives the super-Schwarzian action for the edge mode in the $osp(2|1)$ BF formulation:
\begin{align}
    S_{tot}\,=\,& -\kappa \int d\tau \;\bigg[\big(1- \bth  \bth' \big)\bigg(\sch(u,\tau) +u'^2\bigg) + \frac{1}{2}\big(\bth \bth''' + 3\bth' \bth''\big)\bigg]\cr
    \,=\, & -\kappa \int d\tau d\xi \;\bigg[\ssch[u,\bth;\tau,\xi]+\frac{1}{4}\bth (\bD_{\xi}\bth)^3\bigg]\ . \label{eq: super schwarzian action bf}
\end{align}

For a constant $OSp(2|1)$ group element $\Upsilon_0$ given by
\begin{align}
    \Upsilon_0\,=\, \left[\begin{array}{cc|c}
		d & c & \alpha \\
		b & a & \gamma \\
		\hline 
		\beta & \delta & e
	\end{array} \right] \ , \label{eq: constant osp element}
\end{align}
two $OSp(2|1)$ group elements $g$ and $\Upsilon_0 g$ yield the same gauge field $\mathbf{a}$ via the Maurer-Cartan form in Eq.~\eqref{eq: Maurer-Cartan form}. Hence the Maurer-Cartan form parametrization of $a$ has redundancy in
\begin{align}
    \Upsilon_0 \, g\;\sim \; g \ ,
\end{align}
By the Iwasawa decomposition~\eqref{eq: iwasawa-like decomposition}, this equivalence relation can be translated into the equivalence relation of $u(\tau)$ and $\bth(\tau)$ as follows.
\begin{align}
	\tan {u\over 2} \, \sim\,&       {a \tan {u\over 2} +b  + \gamma \sqrt{{u' \over 2}} \sec {u \over 2} \bth \over c\tan {u\over 2} +d + \alpha \sqrt{{u' \over 2}}  \sec{u \over 2} \bth  } \label{eq: osp gauge bf 1}\\
	\sqrt{{u' \over 2}} \sec{u \over 2}\bth \,\sim \, &      { \delta \tan {u\over 2} + \beta + e \sqrt{{u' \over 2}} \sec{u \over 2} \bth   \over  c \tan {u\over 2} + d+ \alpha \sqrt{{u' \over 2}} \sec{u \over 2} \bth }\label{eq: osp gauge bf 2}
\end{align}
As pointed out in Section~\ref{sec: osp gauging}, this redundancy, or in other words $OSp(2|1)$ gauging of the super-Schwarzian theory, is not identical to the isometry of the $\ads_2$ background. In the BF formulation, this equivalence relation is related to the ``gauge'' transformation which does not change the background gauge field $\mathbf{a}_\ads$ although it is non-trivial\footnote{\ie ``non-trivial'' means that the gauge parameter is not identity on the boundary.} on the boundary.
\begin{align}
    \Upsilon^{-1}\mathbf{a}_\ads\Upsilon^{-1}+ \Upsilon^{-1}\partial_\tau \Upsilon^{-1} \,=\, \mathbf{a}_\ads \ ,
\end{align}
where the gauge parameter $\Upsilon$ is expressed in term of the constant $OSp(2|1)$ group element in Eq.~\eqref{eq: constant osp element} by
\begin{align}
    \Upsilon \,\equiv \, e^{-\mathbf{a}_\ads \tau } \Upsilon_0 e^{\mathbf{a}_\ads \tau } \ .
\end{align}

The path integral measure in the $sl(2,\mathbb{R})$ BF formulation can be obtained from the Haar measure~\cite{Cotler:2018zff,Joung:2023doq}. From the Haar measure of $OSp(2|1)$~\cite{Fan:2021wsb} we obtain the Haar measure for the Iwasawa-like decomposition~\eqref{eq: iwasawa-like decomposition}:
\begin{align}
    \cD\mu_{\text{\tiny OSp Haar}}\,=\, {1\over 8 y^2}\left[ Du \,  Dy \,  Df | D\bth \, D\eta \right]\ .
\end{align}
Using the constraints in the asymptotic AdS condition~\eqref{eq: asymptotic ads condition in bf}, the path integral measure for the edge mode can be writen as
\begin{align}
	D \mu_{\text{\tiny edge}} \,=\, \delta(\mathbf{a}_{11}) \delta(\mathbf{a}_{21}-1) \delta(\mathbf{a}_{31}) D \mu_{\text{\tiny OSp Haar}}
\end{align}
By solving the delta function, the path integral measure for the super-Schwarzian is found to be
\begin{equation}
D \mu_{\text{\tiny edge}} \,=\,  Du D\bth \prod_\tau{1\over 4 u'(\tau)} \ .
\end{equation}

One may use a similar decomposition in which the position of the matrices with fermionic coordinate differs from Eq.~\eqref{eq: iwasawa-like decomposition}. Those different decompositions are related by field redefinitions, and one would expect that the result will be equivalent. For example, we obtain the solutions of the asymptotic AdS condition~\eqref{eq: asymptotic ads condition in bf}, the $OSp(2|1)$ gauging~\eqref{eq: osp gauge bf 1}$\sim$\eqref{eq: osp gauge bf 2} and the action for the edge mode~\eqref{eq: super schwarzian action bf} in other decompositions. And we confirm that the boundary action and the $OSp(2|1)$ gauging are equivalent via a field redefinition. However it still remains open question to prove the equivalence of the path integral measure and the solution in each decomposition.

%%%%%%%%%%%%%%%%%%%%%%%%%%%%%%%%%%%%%%%%%%%%%%%%%%%%%%%%%%%
%%%%%%%%%%%%%%%%%%%%%%%%%%%%%%%%%%%%%%%%%%%%%%%%%%%%%%%%%%%
\section{Conclusion}
\label{sec: conclusion}
%%%%%%%%%%%%%%%%%%%%%%%%%%%%%%%%%%%%%%%%%%%%%%%%%%%%%%%%%%%
%%%%%%%%%%%%%%%%%%%%%%%%%%%%%%%%%%%%%%%%%%%%%%%%%%%%%%%%%%%

We have studied the gravitational edge mode in $\cN=1$ JT supergravity at finite temperature. We reviewed the derivation of gravitational edge modes in bosonic JT gravity, laying the groundwork for our extension into the supersymmetric case. In the context of bosonic JT gravity, we revisited the wiggling boundary description in the conformal gauge. Additionally, we delved into the frame-like formulation of the JT gravity, where we have elaborated on the derivation of the Schwarzian action from the frame field fluctuation. Extending the method used in the bosonic JT gravity, we revisited in $\cN=1$ JT supergravity with the superconformal gauge and an appropriate boundary term. Through both the wiggling boundary description and the superframe fluctuation approach, we derived the super-Schwarzian action for the gravitational edge mode in the $\cN=1$ JT supergravity, highlighting the pivotal role of the supersymmetric version of the inversion formula that maps the base and target super-Poincare disks. The map between base and target super-Poincare disks was also discussed as a method to parameterize these edge modes, aimed at clarifying the origin of the redundancy in their description. This discussion highlighted the subtle yet crucial distinctions between isometry and $OSp(2|1)$ gauging, which supports our inversion procedure for the super-Schwarzian action.

We further examined the gravitational edge mode in the $osp(2|1)$ BF formulation of $\cN=1$ JT supergravity. We discussed the asymptotic AdS condition, treating it as a constraint within the BF formulation framework. Employing the Iwasawa-like decomposition, which naturally captures monodromy associated with disk topology, we derived the super-Schwarzian action. Our analysis highlighted how the inherent redundancy in the Iwasawa-like decomposition gives right to $OSp(2|1)$ gauging of the super-Schwarzian action. Furthermore, we discussed the derivation of the path integral measure from the Haar measure of $OSp(2|1)$ group together with the constraints.

The map between the base and target super-Poincare disk in our analysis is a special example of the immersion discussed in Ref.~\cite{Stanford:2020qhm,Ferrari:2024kpz}. We have focused on the case where the image of the boundary of the base disk is close to the boundary of the target disk, essentially exploring the small wiggling limit. It would be interesting to examine the large wiggling case to uncover the dynamics beyond the super-Schwarzian description. Moreover, the extension of the self-overlapping curve, investigated in Ref.~\cite{Stanford:2020qhm,Ferrari:2024kpz}, to the super-Poincare disk offers another promising avenue to explore.

Expanding our analysis to the $\cN=2$ and $\cN=4$ JT supergravity demands a deeper understanding of mappings not just between two superspaces but also between their internal spaces. This pursuit could unveil richer structures and symmetries inherent in higher supersymmetric extensions of JT supergravity. Understanding these mappings could also pave the way for new insights into the interplay between geometry and quantum gravity, potentially offering novel perspectives on the holographic principle.

\acknowledgments

We would like to thank Imtak Jeon for useful discussions. K.L. was supported by Basic Science Research Program through the National Research Foundation of Korea(NRF) funded by the Ministry of Education(NRF-2020R1I1A2054376). The work of K.L. was supported by a KIAS Individual Grant (PG096201) at Korea Institute for Advanced Study. The work of J.Y. and A.S. was supported by the National Research Foundation of Korea (NRF) grant funded by the Korea government (MSIT) (No. NRF-2022R1A2C1003182). The work of J.Y. and A.S. was supported by the National Research Foundation of Korea (NRF) grant funded by the Korea government (MSIT) (No. NRF-2023K2A9A1A01095488). J.Y. is supported by an appointment to the JRG Program at the APCTP through the Science and Technology Promotion Fund and Lottery Fund of the Korean Government. J.Y. is also supported by the Korean Local Governments - Gyeongsangbuk-do Province and Pohang City.

\appendix

%%%%%%%%%%%%%%%%%%%%%%%%%%%%%%%%%%%%%%%%%%%%%%%%%%%%%%%%%%%
%%%%%%%%%%%%%%%%%%%%%%%%%%%%%%%%%%%%%%%%%%%%%%%%%%%%%%%%%%%
\section{Conventions for supergravity}
\label{app: Conventions for supergravity}
%%%%%%%%%%%%%%%%%%%%%%%%%%%%%%%%%%%%%%%%%%%%%%%%%%%%%%%%%%%
%%%%%%%%%%%%%%%%%%%%%%%%%%%%%%%%%%%%%%%%%%%%%%%%%%%%%%%%%%%

We now discuss our conventions for the $(2|2)$-dimensional super geometries (mostly following \cite{Chamseddine:1991fg,Martinec:1983um}). 

The complex coordinates for curved space are denoted as $w,\wb,\etp,\etm$ and for the associated tangent space as $\bw,\bwb,\bep,\bem$. They are collectively denoted as $x^{M}$ and $x^{A}$ respectively. Note that for complex conjugation of Grassmann numbers, we opt the convention where the order of the product \textit{is not} altered. Therefore the combination $i \etp \etm$ is real valued.  

In terms of the super-frame field $\sfr\indices{_{M}^{A}}$ satisfying
\begin{equation}
    \sfr\indices{_{A}^{M}} \sfr\indices{_{M}^{B}} = \delta\indices{_{A}^{B}} \ , \quad \sfr\indices{_{M}^{A}} \sfr\indices{_{A}^{N}} = \delta\indices{_{M}^{N}} \ ,
\end{equation}
the covariant derivatives on the tangent/flat space are of the form
\begin{equation}
\bDp_{A} = \sfr\indices{_{A}^{M}}  \bDp_{M} \ , \qquad \bDp_{M} \equiv \del_{M} +i \spc_{M} \lgen\ , 
\end{equation}
where $\spc$ is the spin-connection and $\lgen$ is the $U(1)$ generator for local Lorentz transformations. In complex co-ordinates $\lgen$ acts diagonally on both Bosonic and Grassmann (flat) vector components via
\begin{equation}
   \lgen V_{\bw} = - V_{\bw} \  ,\quad \lgen V_{\bwb} = V_{\bwb} \ , \quad \lgen V_{\bep} = - \frac{1}{2}V_{\bep} \ , \quad \lgen V_{\bem} = \frac{1}{2} V_{\bem} \ ,
\end{equation}
and the action on raised indices have the opposite parity. If the Lorentz transformation (which is a rotation in Euclidean signature) acts locally, the spin connection undergoes a gauge transformation
\begin{equation}
 V^{A} \to e^{i\lambda \lgen} V^{A}\ , \qquad \spc_{M} \to \spc_{M} - \del_{M} \lambda \ .
\end{equation}
The storsion $T_{AB}^{C}$ and supercurvature $R_{AB}$ associated with $\bDp_{A}$ are given by
\begin{equation}
	\begin{split}
	\left[\bDp_{A},\bDp_{B} \right\}  & = (-)^{MB} \sfr\indices{_A^{M}}\sfr\indices{_B^{N}} \left[K_{NM} - (-)^{MN} K_{MN}\right] \\
	&\equiv T_{AB}^{C} \bDp_{C} - i R_{AB} \lgen \ ,
 		\end{split}  
\end{equation}
where $K_{NM} \equiv \bDp_{N}\sfr\indices{_M^{C}} \bDp_{C} - i \del_{N} \spc_{M}  \lgen $ ,  and 
\begin{equation}
\bDp_{N}\sfr\indices{_M^{C}} = \del_{N}\sfr\indices{_M^C} + i \spc_{N}\lgen\indices{^{C}_{D}} \sfr\indices{_M^{D}} \ .
\end{equation}
More explicitly we have
\begin{equation}
	\begin{split}
	T^{C}_{AB} &= (-)^{MB}\sfr\indices{_A^{M}}\sfr\indices{_B^{N}} \left[ \bDp_{N}\sfr\indices{_M^{C}} - (-)^{MN} \bDp_{M}\sfr\indices{_N^{C}}   \right] \ , \\
	R_{AB} &= (-)^{MB}\sfr\indices{_A^{M}}\sfr\indices{_B^{N}} \left[ \del_{N}\spc_{M} - (-)^{MN}\del_{M} \spc_{N}\right] \ .
	\end{split}
\end{equation}
When written in complex co-ordinates, the $(2|2)$-dimensional supergeometry is defined by constraining the torsion components such that
\begin{equation}
	\begin{split}
T^{\bw}_{\bep \bep} =2 \ ,\qquad  T^{\bwb}_{\bem \bem} =2 \ , \qquad T_{\text{even even}}^{\text{even}} = T_{\text{odd odd}}^{\text{odd}} = 0 \ . \label{eq: supertorsion constraints}
	\end{split}
\end{equation}
The first constraint among these is inspired from flat space superalgebra. The second and third determine the spin connection $\spc_M$ in terms of the superframe $\sfr\indices{_M^{A}}$.

As mentioned in Section~\ref{sec: super jt gravity}, we work with the superconformal gauge \eqref{eq:sucongauge} following \cite{Martinec:1983um,Chamseddine:1991fg}. Imposing the torsion constraints \eqref{eq: supertorsion constraints} on $\bDp_{A}$ yields the spin connections 
\begin{equation}
\spc_\w = - i \dw \swe  \ , \qquad \spc_{\etp}  = - i \dep \swe \ , 
\end{equation}
for the choice of superconformal factor $e^{\swe}$ . The corresponding solutions for the frame field are as given in \eqref{eq:ConfGaugeSuFrame1} and \eqref{eq:ConfGaugeSuFrame2}.

Note that, torsion and curvature tensors are further constrained due to the Bianchi identities. All the remaining components of the torsion tensor either vanish or get related to the super curvature component $R_{\bep \bem}$. The Bianchi identities also completely determine the components of the super curvature $R_{AB}$. In short, all the non-trivial curvature and torsion quantities can be derived from $R_{\bep \bem}$.

%%%%%%%%%%%%%%%%%%%%%%%%%%%%%%%%%%%%%%%%%%%%%%%%%%%%%%%%%%%
\section{Conventions for \texorpdfstring{$osp(2|1)$}{osp(2|1)} algebra}
\label{app: osp algebra}
%%%%%%%%%%%%%%%%%%%%%%%%%%%%%%%%%%%%%%%%%%%%%%%%%%%%%%%%%%%

The osp$(2|1)$ algebra consists of the bosonic generators $L_{n}$ and fermionic generators $G_{{p}}$, where $m=0, \pm1$ and $p=\pm \frac{1}{2}$. They satisfy
\begin{equation}
	\begin{split}
		\sbk{L_{m},L_{n}} &\,=\, (m-n)L_{m+n}  \ , \quad \sbk{L_{m},G^{I}_{p}} = \left(\frac{1}{2} m -p \right) G_{m+p} \ ,\\
		 \cbk{G_{p},G_{q}} &\,=\, - 2  L_{p+q} \ .
	\end{split}
\end{equation}
An explicit representation can be written as
\begin{equation}
	\begin{split}
		L_{1} \,=\,& \left[ \begin{array}{c c|c}
	0 & 0 & 0 \\
	1 & 0 & 0 \\
	\hline
	0 & 0 & 0 \\
\end{array} \right] \ , \quad L_{0} \,=\,\left[ \begin{array}{cc|c}
			\frac{1}{2} & 0 & 0\\
			0 & - \frac{1}{2} & 0  \\   
			\hline 
			0& 0 & 0   \\
		\end{array}\right] 
	 \ , \quad 
L_{-1} \,=\, \left[ \begin{array}{c c|c}
	0 & -1 & 0 \\
	0 & 0 & 0 \\
	\hline
	0 & 0 & 0 \\
\end{array} \right]  \ ,\\
G_{\frac{1}{2}} \,=\,& \left[ \begin{array}{c c|c}
0 & 0 & 0 \\
0 & 0& 1 \\
\hline
-1 & 0 & 0\\
\end{array}\right] \ , \quad G_{-\frac{1}{2}} \,=\, \left[ \begin{array}{c c|c}
0 & 0 &1 \\
0 & 0& 0 \\
\hline
0 & 1 & 0\\
\end{array}\right] \ .
	\end{split}
\end{equation}
In this basis, the Killing form is given by
\begin{align}
    \str \big(L_nL_m\big)\,=\, \begin{pmatrix}
    0 & 0 & -1\\
    0 & {1\over 2} & 0\\
    -1 & 0 & 0\\
    \end{pmatrix}\ ,\quad \str \big(G_\alpha G_\beta\big)\,=\, \begin{pmatrix}
    0 & 2\\
    -2 & 0\\
    \end{pmatrix}
\end{align}
where the supertrace $\textit{str}$ is defined for a matrix $M$ as
\begin{equation}
 \text{str} M \equiv    M_{11} + M_{22}  - M_{33} \ .
\end{equation}
The basis used in the main text is related to the above via
\begin{equation}\label{eq:JQgen}
	\begin{split}
			J_{0} &\equiv \frac{1}{2} (L_{_+} - L_{_-}) \ , \ J_{1} \equiv L_{0} \ ,\ J_{2} \equiv \frac{1}{2} (L_{_+}+L_{_-}) \ ,  \\
			 Q_{_+} &\equiv \frac{1}{2} G_{_-} \ , \quad  Q_{_-} \equiv \frac{1}{2} G_{_+} \ .
	\end{split}
\end{equation}
They satisfy
\begin{equation}
\begin{split}
	\sbk{J_{I},J_{J}} &=J^{K}\,\tensor{\epsilon}{_{K}_{I}_{J}} \ , \quad  \sbk{J_{I},Q_{\alpha}}= \frac{1}{2} Q_{\beta} \, \left(\Gamma_{I}\right)^{\beta}_{\alpha}, \\
 \cbk{Q_{\alpha},Q_{\beta}} &= -\frac{1}{2} J^{I} \left(C\Gamma _{I}\right)_{\alpha \beta} \ \ .
 \end{split}
\end{equation}
In the above, indices are lowered by right-multiplying with 
\begin{equation}
	\begin{split}
\eta_{I J} &\equiv	2 \, \text{str}\left(J_{I}J_{J} \right) \ , \quad   C_{\alpha \beta}  \equiv 2 \,  \text{str}\left(Q_{\alpha}Q_{\beta}\right)  \ , \\
\eta_{IJ} &= \text{diag}(1,1,-1) \ , \quad C_{_{+-}}= - C_{_{-+}}=-1 \ ,
	\end{split}
\end{equation}
$\epsilon$ is the Levi-Civita tensor with $\epsilon^{012}=1$ and $\Gamma_{I}$ satisfy the Clifford algebra with $\{\Gamma_{I},\Gamma_{J} \} =  2 \eta_{IJ}$. An explicit realization for $\Gamma_{I}$ is given by

\begin{equation}
	\Gamma_{0} = \left[  \begin{array}{c c}
	0 & 1  \\
	1 & 0
	\end{array} \right] \ , \ 
	\Gamma_{1} = \left[  \begin{array}{c c}
	 1 & 0  \\
	0 & -1
\end{array} \right] \ , \ 
\Gamma_{2} = \left[  \begin{array}{c c}
	0  & 1  \\
	-1  & 0 \\
\end{array} \right] \ .
\end{equation}

%%%%%%%%%%%%%%%%%%%%%%%%%%%%%%%%%%%%%%%%%%%%%%%%%%%%%%%%%%%
%%%%%%%%%%%%%%%%%%%%%%%%%%%%%%%%%%%%%%%%%%%%%%%%%%%%%%%%%%%
\section{Asymptotic AdS Condition for \texorpdfstring{$osp(2|1)$}{osp(2|1)} BF formulation}
\label{app: constraint in bf theory}
%%%%%%%%%%%%%%%%%%%%%%%%%%%%%%%%%%%%%%%%%%%%%%%%%%%%%%%%%%%
%%%%%%%%%%%%%%%%%%%%%%%%%%%%%%%%%%%%%%%%%%%%%%%%%%%%%%%%%%

In this appendix, we illustrate how the asymptotic AdS condition in Eq.~\eqref{eq: asymptotic ads condition in bf} can be understood from the perspective of constraints.

For the case of the $sl(2,\mathbb{R})$ BF theory, the asymptotic AdS condition was incorporated by the nilpotent subgroup~\cite{Valach:2019jzv,Joung:2023doq}. In the way, we introduce the nilpotent subgroup element $c$ in $OSp(2|1)$ via gauge transformation of $\mathbf{a}$:
\begin{align}
    \mathbf{a}\quad\longrightarrow \quad c^{-1} \mathbf{a} c +  c^{-1}\partial_\tau c\ ,\label{eq: transf by nilpotent}
\end{align}
where the gauge field $\mathbf{a}$ and the nilpotent subgroup element $c$ are given by
\begin{align}
    \mathbf{a}(\tau)\,=\,  \left[  \begin{array}{c c| c}
    {1\over 2}a_0 & a_{-1} & \psi_{-{1\over 2}}\\
    a_1 & -{1\over 2}a_0 & \psi_{1\over 2} \\
    \hline
    -\psi_{1\over 2} & \psi_{-{1\over 2}} & 0\\
    \end{array}\right]\;\;,\quad c(\tau)\,=\, \left[  \begin{array}{c c| c}
    1 & \lambda & \zeta \\
    0 &1 & 0 \\
    \hline
    0 & \zeta & 1\\
    \end{array}\right]\ .
\end{align}
The bulk action~\eqref{eq: bulk action bf} vanishes because we solved the bulk constraint by flat connection while the boundary action is not invariant under the ``gauge'' transformation by the nilpotent element. Therefore, the total action becomes
\begin{align}
    S_{tot}\,=\, &{\kappa\over 2}\int d\tau\, \str(\mathbf{a}^2)\ ,\cr
    \longrightarrow \quad S_{tot}\,=\, &{\kappa\over 2} \int d\tau \, \str \big(\mathbf{a}^2 +2 \mathbf{a}\,  \partial_\tau c c^{-1} \big)\ ,\cr
    \,=\,&\kappa \int d\tau \bigg( {1\over 4}a_0^2- a_1 a_{-1} + a_1 \partial_\tau \lambda + 2\psi_{1\over 2} \psi_{-{1\over 2}}+ 2\psi_{1\over 2}\partial_\tau \zeta + a_1 \zeta \partial_\tau \zeta\bigg)\ .
\end{align}
We promote the nilpotent element $c$ to a dynamical field like Stuekelberg field. Then by construction, we recover the gauge symmetry amount to the nilpotent subgroup:
\begin{equation}
\begin{split}
    \mathbf{a}\quad &\longrightarrow\quad k^{-1}\mathbf{a}k +k^{-1}\partial_\tau k\\
    c\quad&\longrightarrow\quad k^{-1}c
\end{split}\ ,\label{eq: gauge transf nilpot bf}
\end{equation}
where the gauge parameter $k$ is given by
\begin{align}
    k(\tau)\,=\, \begin{pmatrix}
    1 & \gamma & \eta \\
    0 & 1 & 0 \\
    0 & \eta & 1\\
    \end{pmatrix}\ .
\end{align}
Note that the new fermionic field $\zeta$ is analogous to the extra fermion introduced to the constrained $OSp(2|1)$ WZW model~\cite{Bershadsky:1989tc}.

By integrating out $\lambda$, one can impose the constraint $a_1=1$, and the nilpotent gauge transformation in Eq.~\eqref{eq: gauge transf nilpot bf} becomes
\begin{align}
    \mathbf{a}\;\;&\longrightarrow\;\; k^{-1}\mathbf{a}k +k^{-1}\partial_\tau k\,=\,\left[  \begin{array}{c c| c}
    {1\over 2}a_0 - \psi_{1\over 2} \eta- \gamma & \ast & \;\;\ast\;\; \\
    1 & -{1\over 2}a_0 + \psi_{1\over 2} \eta+ \gamma &\;\;\ast\;\;  \\
    \hline
    -\psi_{1\over 2} - \eta &\ast  & 0 \\
   \end{array}\right] \ ,\\
    h\;\;&\longrightarrow\;\; k^{-1}h\,=\,\left[  \begin{array}{c c| c}
    1 & \lambda-\gamma-\eta \zeta & -\eta +\zeta\\
    0 & 1 & 0\\
    \hline
    0 & \zeta-\eta & 1\\
    \end{array}\right] \ .
\end{align}
Note that $\psi_{1\over 2}+\zeta$ is invariant under the nilpotent gauge transformation in Eq.~\eqref{eq: gauge transf nilpot bf}. Hence we can impose the following constraint which is invariant under the above gauge transformation~\cite{Bershadsky:1989tc}.
\begin{align}
    a_1\,=\, 1\;\;,\quad \psi_{1\over 2}+\zeta\,=\,0\ .
\end{align}
Furthermore, by using the gauge transformation~\eqref{eq: gauge transf nilpot bf}, we can choose the gauge condition:
\begin{align}
    \zeta\,=\,0\quad,\qquad a_0\,=\,0\ .
\end{align}
Finally the action is found to be
%
%\begin{align}
%    S_{tot}\,=\,& \int d\tau \bigg({1\over 4}a_0^2- a_{-1} +   2\psi_{1\over 2}\psi_{-{1\over 2}}+ 2\psi_{1\over 2}\partial_\tau \zeta + \zeta \partial_\tau \zeta\bigg)
%\end{align}
%
%\begin{align}
%    S_{tot}\,=\,\int d\tau \,\bigg( {1\over 4}a_0^2- a_{-1} - 2\zeta\psi_{-{1\over 2}}- \zeta\partial \zeta \bigg)
%\end{align}
%
\begin{align}
    S_{tot}\,=\,-\kappa \int d\tau \, a_{-1}\ ,
\end{align}
with the asymptotic AdS condition
\begin{align}
    \mathbf{a}\,=\,   \left[  \begin{array}{c c| c}
    0 & a_{-1} & \psi_{-{1\over 2}}\\
    1 & 0 & 0 \\
    \hline
    0 & \psi_{-{1\over 2}} & 0\\
    \end{array}\right] \ .
\end{align}

%%%%%%%%%%%%%%%%%%%%%%%%%%%%%%%%%%%%%%%%%%%%%%%%%%%%%%%%%%%
%%%%%%%%%%%%%%%%%%%%%%%%%%%%%%%%%%%%%%%%%%%%%%%%%%%%%%%%%%%
\section{Different Iwasawa-like Decomposition and Path Integral Measure}
\label{app: path integral measure bf}
%%%%%%%%%%%%%%%%%%%%%%%%%%%%%%%%%%%%%%%%%%%%%%%%%%%%%%%%%%%
%%%%%%%%%%%%%%%%%%%%%%%%%%%%%%%%%%%%%%%%%%%%%%%%%%%%%%%%%%

In this appendix, we present the other Iwasawa-like decompositions. The maximal compact group element is always located in the decomposition at the leftmost while the position of the nilpotent element with fermionic coordinates differs in each decomposition.

\paragraph{Parameterization 1 (Section~\ref{sec: bf description}):} 

\begin{itemize}
\item Iwasawa-like Decomposition 1
\begin{align}
		g_1\, \equiv \,&  \left[\begin{array}{c c | c}
			\cos \frac{ u_1(\tau)}{2} & -\sin \frac{ u_1(\tau)}{2}  & 0\\
			\sin \frac{ u_1(\tau)}{2} & \cos \frac{ u_1(\tau)}{2} & 0 \\
			\hline
			0 & 0 & 1 \\
		\end{array}\right]  \left[ \begin{array}{c c|c}
	 [y_1(\tau)]^{-{1\over 2}} & 0 & 0\\
	 0 & [y_1(\tau)]^{1\over 2} & 0\\
	 \hline
	 0 & 0 & 1
     \end{array}\right] \cr
     &\times\left[\begin{array}{c c | c}
  	                      1 & 0 & 0 \\
  	                      0 & 1 & - \bth_1(\tau)\\
  	                      \hline
  	                      \bth_1(\tau) & 0 & 1
  \end{array} \right]\left[\begin{array}{c c | c}
  1 & 0 & \eta_1(\tau) \\
  0 & 1 & 0\\
  \hline
  0 & \eta_1(\tau) & 1
\end{array} \right] \left[ \begin{array}{c c|c}
	 1 & - f_1(\tau) & 0\\
	 0 & 1 & 0\\
	 \hline
	 0 & 0 & 1
     \end{array}\right]\ .\label{eq: iwasawa-like decomposition 1}
\end{align}
\item Solution of the asymptotic AdS condition
\begin{align}
    y_1\,=\,&  \frac{1}{2} u'_1\big(1+ \bth_1  \bth_1' \big)\ ,\\
    f_1\,=\,&   \frac{1}{2} \bigg({u_1''\over u_1'}+\bth_1 \bth_1''\bigg)\ ,\\
    \eta_1\,=\,& \bth_1' + \frac{1}{2}\bth_1  {u_1''\over u_1'}\ ,
\end{align}
\begin{align}
    \mathcal{L}_1\,=\,& \frac{1}{2} \big(1- \bth_1  \bth_1' \big)\bigg(\sch(u_1,\tau) +\frac{1}{2} u_1'^2\bigg) + \frac{1}{2}\big(\bth_1 \bth_1''' + 3\bth_1' \bth_1''\big) \ ,\\
    \mathcal{G}_1\,=\,& \frac{1}{2} \bth_1 \bigg(\sch(u_1,\tau) +\frac{1}{2} u_1'^2 \bigg) + \bth_1'' + \frac{1}{2}\bth_1  \bth_1'  \bth_1''\ .
\end{align}
\item $OSp(2|1)$ gauging (redundancy)
\begin{align}
	\tan {u_1\over 2} \, \sim\,&       {a \tan {u_1\over 2} +b  + \gamma \sqrt{{u_1' \over 2}} \sec {u_1 \over 2} \bth_1 \over c\tan {u_1\over 2} +d + \alpha \sqrt{{u_1' \over 2}}  \sec{u_1 \over 2} \bth_1  }\ , \\
	\sqrt{{u_1' \over 2}} \sec{u_1 \over 2}\bth_1 \,\sim \, &      { \delta \tan {u_1\over 2} + \beta + e \sqrt{{u_1' \over 2}} \sec{u_1 \over 2} \bth_1   \over  c \tan {u_1\over 2} + d+ \alpha \sqrt{{u_1' \over 2}} \sec{u_1 \over 2} \bth_1 }\ ,
\end{align}
where
\begin{align}
    \Upsilon_0\,=\, \left[\begin{array}{cc|c}
		d & c & \alpha \\
		b & a & \gamma \\
		\hline 
		\beta & \delta & e
	\end{array} \right] \,\in \, OSp(2|1)\ .
\end{align}
\item Path integral measure from Haar measure
\begin{equation}
D \mu^1_{\text{\tiny edge}} =  Du_1 D\bth_1 \prod_\tau{1\over 4 u'(\tau_1)} \ .
\end{equation}

\end{itemize}

\paragraph{Parameterization 2:} 
\begin{itemize}
\item Iwasawa-like Decomposition 2
\begin{align}
		g_2\, \equiv \,&  \left[\begin{array}{c c | c}
			\cos \frac{ u_2(\tau)}{2} & -\sin \frac{ u_2(\tau)}{2}  & 0\\
			\sin \frac{ u_2(\tau)}{2} & \cos \frac{ u_2(\tau)}{2} & 0 \\
			\hline
			0 & 0 & 1 \\
		\end{array}\right] 
     \left[\begin{array}{c c | c}
  	                      1 & 0 & 0 \\
  	                      0 & 1 & - \bth_2(\tau)\\
  	                      \hline
  	                      \bth_2(\tau) & 0 & 1
  \end{array} \right]\left[\begin{array}{c c | c}
  1 & 0 & \eta_2(\tau) \\
  0 & 1 & 0\\
  \hline
  0 & \eta_2(\tau) & 1
\end{array} \right]\cr  
    &\times \left[ \begin{array}{c c|c}
	 [y_2(\tau)]^{-{1\over 2}} & 0 & 0\\
	 0 & [y_2(\tau)]^{1\over 2} & 0\\
	 \hline
	 0 & 0 & 1
     \end{array}\right] \left[ \begin{array}{c c|c}
	 1 & - f_2(\tau) & 0\\
	 0 & 1 & 0\\
	 \hline
	 0 & 0 & 1
     \end{array}\right]\ ,\label{eq: iwasawa-like decomposition 2}
\end{align}
\item Solution of the asymptotic AdS condition
\begin{align}
    y_2\,=\,&  \frac{1}{2} u_{2}' + \bth_{2}\bth_{2}'\ ,\\
    f_2\,=\,&   \frac{1}{2} \frac{u_{2}''}{u_{2}'}+\bth_{2} \left(\frac{\bth_{2}'}{u_{2}'} \right)^{_{'}}\ ,\\
    \eta_2\,=\,& 2 \frac{\bth_{2}'}{u_{2}'}\ ,
\end{align}
\begin{align}
    \mathcal{L}_2\,=\,&  \left(\frac{1}{2}- \frac{\bth_2  \bth_2'}{u_{2}'} \right)\bigg(\sch(u_2,\tau) +\frac{1}{2} u_2'^2\bigg) + \frac{1}{u_{2}'}\big(\bth_2 \bth_2''' + 3\bth_2' \bth_2''\big) - \frac{3u_{2}''}{(u_{2}')^{3 \over 2}}\left(\frac{\bth_{2}\bth_{2}'}{\sqrt{u_{2}'}} \right)^{_{'}} \ ,\\
    \mathcal{G}_2\,=\,& \sqrt{2 u_{2}'} \left[\left(\frac{\bth_{2}'}{u_{2}'} \right)^{_{'}}+ \frac{1}{4} u_{2}'\bth_{2} + \frac{1}{(u_{2}')^{2}} \bth_{2}\bth_{2}'\bth_{2}'' \right]\ .
\end{align}
\item $OSp(2|1)$ gauging (redundancy)
\begin{align}
	\tan {u_2\over 2} \, \sim\,&       {a \tan {u_2\over 2} +b  +  \gamma \sec {u_2 \over 2} \bth_2 \over c\tan {u_2\over 2} +d + \alpha   \sec{u_2 \over 2} \bth_2  } \\
	 \sec{u_2 \over 2}\bth_2 \,\sim \, &      { \delta \tan {u_2\over 2} + \beta + e  \sec{u_2 \over 2} \bth_2   \over  c \tan {u_2\over 2} + d+ \alpha  \sec{u_2 \over 2} \bth_2 }
\end{align}
where
\begin{align}
    \Upsilon_0\,=\, \left[\begin{array}{cc|c}
		d & c & \alpha \\
		b & a & \gamma \\
		\hline 
		\beta & \delta & e
	\end{array} \right] \,\in \, OSp(2|1) \ .
\end{align}
\item Path integral measure from Haar measure
\begin{equation}
D \mu^2_{\text{\tiny edge}} =  Du_2 D\bth_2 \prod_\tau{1- \frac{\bth_{2}(\tau)\bth_{2}'(\tau)}{u_{2}'(\tau)}  \over 4\sqrt{2} \sqrt{u_2'(\tau)}} \ .
\end{equation}
\end{itemize}

\paragraph{Parameterization 3:} 
\begin{itemize}
\item Iwasawa-like Decomposition 2
\begin{align}
		g_3\, \equiv \,&  \left[\begin{array}{c c | c}
			\cos \frac{ u_3(\tau)}{2} & -\sin \frac{ u_3(\tau)}{2}  & 0\\
			\sin \frac{ u_3(\tau)}{2} & \cos \frac{ u_3(\tau)}{2} & 0 \\
			\hline
			0 & 0 & 1 \\
		\end{array}\right] \left[\begin{array}{c c | c}
  	                      1 & 0 & 0 \\
  	                      0 & 1 & - \bth_3(\tau)\\
  	                      \hline
  	                      \bth_3(\tau) & 0 & 1
  \end{array} \right] \left[ \begin{array}{c c|c}
	 [y_3(\tau)]^{-{1\over 2}} & 0 & 0\\
	 0 & [y_3(\tau)]^{1\over 2} & 0\\
	 \hline
	 0 & 0 & 1
     \end{array}\right] \cr
     &\times\left[\begin{array}{c c | c}
  1 & 0 & \eta_3(\tau) \\
  0 & 1 & 0\\
  \hline
  0 & \eta_3(\tau) & 1
\end{array} \right] \left[ \begin{array}{c c|c}
	 1 & - f_3(\tau) & 0\\
	 0 & 1 & 0\\
	 \hline
	 0 & 0 & 1
     \end{array}\right]\ ,\label{eq: iwasawa-like decomposition 3}
\end{align}
\item Solution of the asymptotic AdS condition
\begin{align}
    y_3\,=\,&  \frac{1}{2} u_{3}' + \bth_{3} \bth_{3}'\ ,\\
    f_3\,=\,&   \frac{1}{2} \bigg({u_3''\over u_3'}+\bth_3 \bth_3''\bigg)\ ,\\
    \eta_3\,=\,& \sqrt{2 \over u_{3}'} \frac{\bth_{3}'}{u_{3}'}\left(u_{3}'+\bth_{3}\bth_{3}' \right)\ ,
\end{align}
\begin{align}
    \mathcal{L}_3\,=\,&  \left(\frac{1}{2}- \frac{\bth_3  \bth_3'}{u_{3}'} \right)\bigg(\sch(u_3,\tau) +\frac{1}{2} u_3'^2\bigg) + \frac{1}{u_{3}'}\big(\bth_3 \bth_3''' + 3\bth_3' \bth_3''\big)- \frac{3 u_{3}''}{\big(u_{3}' \big)^{\frac{3}{2}}} \left(\frac{\bth_{3}\bth_{3}'}{\sqrt{u_{3}'}} \right)^{_{'}} \ ,\\
    \mathcal{G}_3\,=\,& \sqrt{2 u_{3}'}\left[ \left(\frac{\bth_{3}'}{u_{3}'} \right)^{_{'}}+ \frac{1}{4}\bth_{3}'u_{3}' + \frac{\bth_{3}\bth_{3}'\bth_{3}''}{\big(u_{3}'\big)^{2}}\right]\ .
\end{align}
\item $OSp(2|1)$ gauging (redundancy)
\begin{align}
	\tan {u_3\over 2} \, \sim\,&       {a \tan {u_3\over 2} +b  + \gamma  \sec {u_3 \over 2} \bth_3 \over c\tan {u_3\over 2} +d + \alpha   \sec{u_3 \over 2} \bth_3  } \\
	 \sec{u_3 \over 2}\bth_3 \,\sim \, &      { \delta \tan {u_3\over 2} + \beta + e  \sec{u_3 \over 2} \bth_3   \over  c \tan {u_3\over 2} + d+ \alpha  \sec{u_3 \over 2} \bth }
\end{align}
where
\begin{align}
    \Upsilon_0\,=\, \left[\begin{array}{cc|c}
		d & c & \alpha \\
		b & a & \gamma \\
		\hline 
		\beta & \delta & e
	\end{array} \right] \,\in \, OSp(2|1) \ .
\end{align}
\item Path integral measure from Haar measure
\begin{equation}
D \mu^3_{\text{\tiny edge}} =  Du_3 D\bth_3 \prod_\tau{ \left(1-\frac{2}{u_{3}'}\bth_{3}\bth_{3}' \right)\over 4 u_3'(\tau)} \ .
\end{equation}

\end{itemize}

\paragraph{From parameterization 2 to Parameterization 1:}

Results for the parameterization 1 can be retrieved from parameterization 2 via the field redefinition
\begin{equation}
(u_{2},\bth_{2}) \to (u_{1},\sqrt{\frac{u_{1}}{2}'}\bth_{1})
\end{equation}

\paragraph{From parameterization 3 to Parameterization 1:}

Results for the parameterization 1 can be retrieved from parameterization 2 via the field redefinition
\begin{equation}
(u_{3},\bth_{3}) \to (u_{1},\sqrt{\frac{u_{1}}{2}'}\bth_{1})
\end{equation}

%%%%%%%%%%%%%%%%%%%%%%%%%%%%%%%%%%%%%%%%%%%%%%%
% \bibliographystyle{JHEP}
\bibliographystyle{JHEP}
\bibliography{superjt}

\end{document}